\pdfoutput=1
\documentclass[letterpaper,twocolumn]{aastex6}

\usepackage{graphicx}
\usepackage{natbib}
\usepackage{amsmath}
\newcommand       \be		{\begin{equation}}
\newcommand       \ee		{\end{equation}}

\newcommand       \kpc		{\,{\rm kpc \,}}
\newcommand       \pc		{\,{\rm pc \,}}
\newcommand       \erg		{\,{\rm erg \,}}

\newcommand       \Myr		{\,{\rm Myr \,}}
\newcommand       \Myrs     {\,{\rm Myrs \,}}

\newcommand       \s		{\,{\rm s \,}}

\newcommand       \kms      {\,{\rm km \,\, s}^{-1}}

\begin{document}

\title{Observational Evidence of Dynamic Star Formation Rate\\ in Milky Way Giant Molecular Clouds}
\author{Eve J. Lee\altaffilmark{1}, Marc-Antoine Miville-Desch\^{e}nes\altaffilmark{2}, Norman W. Murray\altaffilmark{3,4}}
\altaffiltext{1}{Astronomy Department, University of California, Berkeley, CA 94720, USA; evelee@berkeley.edu}
\altaffiltext{2}{Institut d'Astrophysique Spatiale, CNRS/Universit\'{e} Paris-Sud 11, 91405 Orsay, France}
\altaffiltext{3}{Canadian Institute for Theoretical Astrophysics, 60 St. George Street, University of Toronto, Toronto ON M5S 3H8, Canada}
\altaffiltext{4}{Canada Research Chair in Theoretical Astrophysics}

\begin{abstract}
Star formation on galactic scales is known to be a slow process, but
whether it is slow on smaller scales is uncertain. 
We cross-correlate 5469 giant molecular clouds (GMCs) 
from a new all-sky catalog with 256 star forming
complexes (SFCs) to build a sample of 
191 SFC-GMC complexes---collections of multiple clouds each matched to 191 SFCs.
The total mass in stars harbored by these clouds 
is inferred from {\it WMAP} free-free fluxes.
We measure the GMC mass, the virial parameter, the star formation efficiency $\epsilon$ 
and the star formation rate per free-fall time $\epsilon_{\rm ff}$. Both $\epsilon$
and $\epsilon_{\rm ff}$ range over 3--4 orders of magnitude.
We find that 68.3\% of the clouds 
fall within $\sigma_{\log\epsilon}=0.79\pm0.22\,{\rm
dex}$ and $\sigma_{\log\epsilon_{\rm ff}}=0.91\pm0.22\,{\rm
dex}$ about the median. Compared to these observed scatters, 
a simple model 
with a time independent $\epsilon_{\rm ff}$ that
depends on the host
GMC properties predicts 
$\sigma_{\log\epsilon_{\rm ff}}=0.12$--0.24.
Allowing for a time-variable $\epsilon_{\rm ff}$, we can recover the
large dispersion in the rate of star formation. This
strongly suggests that star formation in the Milky Way is a 
dynamic process on GMC scales. We also show that the surface star
formation rate profile of the Milky Way correlates well 
with the molecular gas surface density profile.
\end{abstract}

\section{Introduction}
\label{sec:intro}

Star formation is a slow process on galactic size and time scales, 
with a mere $\sim 2\%$ of the gas mass turning into stars in the disk dynamical time
\citep{kennicutt89, kennicutt98}. Stars
in the Milky Way and nearby galaxies form in giant molecular clouds (GMCs),
with the mass in newly formed stars proportional to the mass in host GMCs \citep[see e.g.,][]{mooney88,scoville89}. 
The bulk of the molecular gas resides in the most massive 
GMCs \citep[e.g.,][]{solomon87} and, as implied by the results of \citet{mooney88}, most 
star formation occurs in the most massive GMCs \citep[see also e.g.,][]{murray11}.

There is considerable disagreement regarding the rate of star formation and the 
star formation efficiencies on scales of GMCs and smaller. 
Star formation efficiency $\epsilon$ is defined as the ratio of the mass in 
protostars to the total mass in a given star-forming region:
\be \label{eqn:efficiency}
\epsilon \equiv \frac{M_\star}{M_g+M_\star},
\ee
where the star-forming region may be a GMC or a smaller sub-region of a GMC.
The star formation rate per free fall time $\epsilon_{\rm ff}$ is
\be \label{eqn:sfe_ff}
\epsilon_{\rm ff}\equiv \epsilon \frac{\tau_{\rm ff}}{\tau_\star},
\ee
where $\tau_*$ is the lifetime of the (proto-)stellar object in question, and 
$\tau_{\rm ff}\equiv\sqrt{3\pi/32G\rho}$ is the free fall time 
of the star-forming region (a GMC or its sub-region), 
which is assumed to have a mean density $\rho$. 

\citet{mooney88} and \citet{scoville89} showed that there is  
a wide (maximum to minimum of approximately $2.5$ dex) 
spread in the efficiencies of star formation in GMCs, a measurement based on 
the ratio of far infrared luminosity $L_{\rm FIR}$ to CO luminosity $L_{\rm CO}$.
More recent estimates also employ
counts of protostars in nearby Milky Way molecular clouds
\citep[e.g.][]{evans09, heiderman10, lada10}, or measurements of 
the free-free emission associated with massive stars \citep{murray11} 
to find a similarly large spread.
\citet{heiderman10} and \citet{lada10} (as well as \citealt{evans09} to a 
smaller degree) find broad distributions
of $\epsilon_{\rm ff}$, which range both well below and 
well above $\epsilon_{\rm ff}=0.02$, by factors of $\sim 20$ or 
more in either direction; they also note that Galactic clouds 
with high surface densities may have 
higher-than-expected $\epsilon_{\rm ff}$ compared to 
their extragalactic counterparts.

\citet{kdm12} argue that the broad distribution 
in $\epsilon_{\rm ff}$ of Galactic molecular clouds can be 
explained by variations in volumetric 
densities (the density determining the free-fall time) among 
clouds. They argue that the star formation rate
on all scales is $\epsilon_{\rm ff}\approx 0.02 M_{\rm GMC}/\tau_{\rm ff}$, 
with only a factor of 3 scatter above and below the mean value,
once the variations in $\tau_{\rm ff}$ are taken into account
\citep[see also e.g.,][]{krumholz07}. 

However, using a sample of clouds from c2d and Gould Belt {\it Spitzer} legacy programs,
\citet{evans14} find a large scatter in $\epsilon_{\rm ff}$ 
even after taking into account the variations in volumetric densities. 
Similarly, \citet{heyer16} study dense clumps in 
ATLASGAL survey and find $\epsilon_{\rm ff} \sim 0.001$--0.01 (see their Figure 10c).

The rate of small scale star formation, and the dispersion of its distribution, 
is important for galaxy formation. If the small scale star formation rate is steady
and hence low, the mass in live stars will be steady, and small.\footnote{We
refer to stars less than 4 Myrs old as ``young" or ``live" stars. Clusters 
containing more than $\sim5000 M_\odot$--$10000 M_\odot$, 
which sample the initial mass function (IMF) 
fairly well, have ionizing 
photon rates $Q\,(\s^{-1})$ that are dominated by massive stars with $M_\star\gtrsim30M_\odot$. Such 
stars have lifetimes of order 4 Myrs.}
It follows that the kinetic and thermal feedback from stellar winds, 
radiation, and supernovae will be steady and low. 
If the small scale star formation is sporadic, with pronounced peaks and long-lived lows, 
the feedback will be both temporally and spatially concentrated. Because stars form
preferentially in dense gas, the resultant feedback will be deposited in regions of 
dense gas, where it has the potential to move the most material, but also where cooling is 
most rapid. Observations of starburst galaxies show that massive star clusters are 
prominent sources of galaxy scale winds \citep[e.g.,][]{schwartz06}. These winds
are believed to be crucial for determining both the global star formation rate and the 
total stellar mass by regulating the amount of gas in the disk \citep[e.g.,][]{oppenheimer10,hopkins11,hopkins14}, 
as well as the distribution of metals in the intergalactic medium \citep[e.g.,][]{oppenheimer10}. 
For a fixed global star formation rate, sporadic small scale star formation will tend 
to produce more massive clusters than will steady small scale star formation, with 
important consequences for wind properties.

Star formation is promoted by gravity and by convergent fluid flows, and
suppressed by a number of physical effects including thermal gas pressure, 
turbulent kinetic pressure, magnetic fields, and stellar feedback --- i.e. stellar winds, 
radiation pressure, protostellar jets, and, at late times, supernovae. 

Two leading candidates (at the time of writing) for the suppression 
of star formation rates are stellar feedback and turbulent pressure support. 
On galactic scales, the rate of star formation is believed to 
be regulated by stellar feedback, which can keep the gas disk in 
a state of marginal stability \citep[e.g.,][]{TQM05, Ostriker_Shetty11}. 
The turbulent pressure scenario
is well motivated: the large linewidths seen in massive star forming regions 
\citep[e.g.,][]{Caselli_Myers95,Plume97} 
show that the kinetic energy density greatly exceeds the thermal pressure on scales larger
than $\sim 0.01\pc$, and is comparable to the gravitational potential 
energy density.\footnote{The two pictures --- stellar feedback 
and turbulence --- are not necessarily 
in conflict, since a source of energy is needed to power the turbulence 
seen in GMCs, and stellar feedback may provide this energy.}
A number of authors have suggested that these turbulent motions 
support GMCs and hence slow the rate of star
formation. The extreme version of the argument says that turbulence maintains GMCs 
in hydrostatic equilibrium, preventing large scale collapse \citep[e.g.,][]{Myers_Fuller_92,mclaughlin97,mckee03}. 

If GMCs are in hydrostatic
equilibrium (which implies that the clouds live for at least a few free fall times) 
then $\epsilon_{\rm ff}= 0.02$ implies that the mass in live stars is roughly the
same in most GMCs of a given mass. However, recent numerical and semi-analytic studies show 
that $\epsilon_{\rm ff}$ increases roughly linearly with time
\citep[e.g.,][]{lee15,murray15,dmurray15}. 
If this is true, then most $10^6M_\odot$ GMCs will have 
very few live stars, while a small subset of $10^6M_\odot$ GMCs will host very 
massive clusters of live stars; the distribution of $\epsilon_{\rm ff}$ 
will be very broad. 

Measuring the width of the distribution in $\epsilon_{\rm ff}$ and $\epsilon$ 
therefore provides an important diagnostic for testing the idea that
$\epsilon_{\rm ff}$ is independent of time. 
We will show that the observed scatter in $\epsilon_{\rm ff}$ of 
the Milky Way GMCs is significantly larger than what is predicted by models of 
constant star formation rate per free fall time~\citep[e.g.,][]{krumholz05,padoan12}.

To estimate either $\epsilon$ or $\epsilon_{\rm ff}$ we must estimate 
the mass in young stars. We use the free-free flux to do so.
Our sample of star-forming Milky Way GMCs is built by cross-correlating 
a new all-sky cloud catalog from M-A., Miville-Desch\^enes et al. 
(2016, in preparation; MML16 from hereon) with star forming complexes 
(SFCs) from \citet{lee12}.

This paper is organized as follows: in Section \ref{sec:sfgmc}, we match 
SFCs with GMCs in (l,b,v)-space (galactic longitude, galactic latitude, radial velocity);
in Section \ref{sec:flux}, we describe how we convert free-free flux into stellar mass; 
in Sections \ref{sec:sfe} and \ref{sec:sfr_ff}, we present our 
analysis of $\epsilon$ and $\epsilon_{\rm ff}$, respectively;
in Section \ref{sec:cause}, we compare 
various models of star formation rate to observations; in Section \ref{sec:sigma}, 
we present the surface density star formation rate profile across the Galactic
plane; we summarize and discuss our results in Section \ref{sec:disc}.

\section{Star Forming Giant Molecular Clouds}
\label{sec:sfgmc}
In this section, we describe how we cross-correlate the SFCs 
in \citet{lee12} with the GMC catalog of MML16.

\subsection{Star Forming Complexes}
\label{ssec:sfc}
We present a brief description of the SFC catalog here \citep[see][for more detail]{lee12}. 
\citet{lee12}, following the approach of \citet{murray10} and \citet{rahman10}, identified 
280 SFCs in and near the Galactic plane. 

Ionizing photons from young clusters can travel 
tens or even hundreds of parsecs through the interstellar medium before being absorbed. 
It follows, and it is observed, that free-free emission regions can be much larger 
($\sim 1$--2$^\circ$ in diameter)
than the star clusters that power them.
In addition, star clusters tend to form in associations (just like 
stars tend to form in clusters), 
so that the free-free emission seen
in a given direction may be powered by more than one star cluster. 
Star forming complexes represent systems of clusters that are able to carve out 
bubbles of size $\sim$10--100 parsecs in their host GMCs.
We assign a free-free flux $f_{\nu}$ to each SFC by dividing up the 
free-free flux of their host {\it WMAP}
sources (the 1--2$^o$ wide regions of free-free emission) in 
proportion to the 8$\mu$m flux of each SFC, as seen by {\it Spitzer}. 
This is motivated by the linear correlation
between free-free and 8$\mu$m flux seen on small (parsec) scales. 

\citet{lee12} calculated the distance to each SFC by fitting the galactic longitude 
and the central local standard at rest velocity $v_{\rm lsr}$ to the \citet[hereafter C85]{clemens85} 
rotation curve. For SFCs that reside inside the solar circle ($R_\odot\leq8.5\kpc$), 
they use the radial velocities of the absorption lines along the line of sight 
to disambiguate between near (absorption line velocity is less than that of 
the SFC) and far (absorption line velocity can be as large as the tangent point 
velocity) distances.

\begin{figure}
\includegraphics[width=0.5\textwidth]{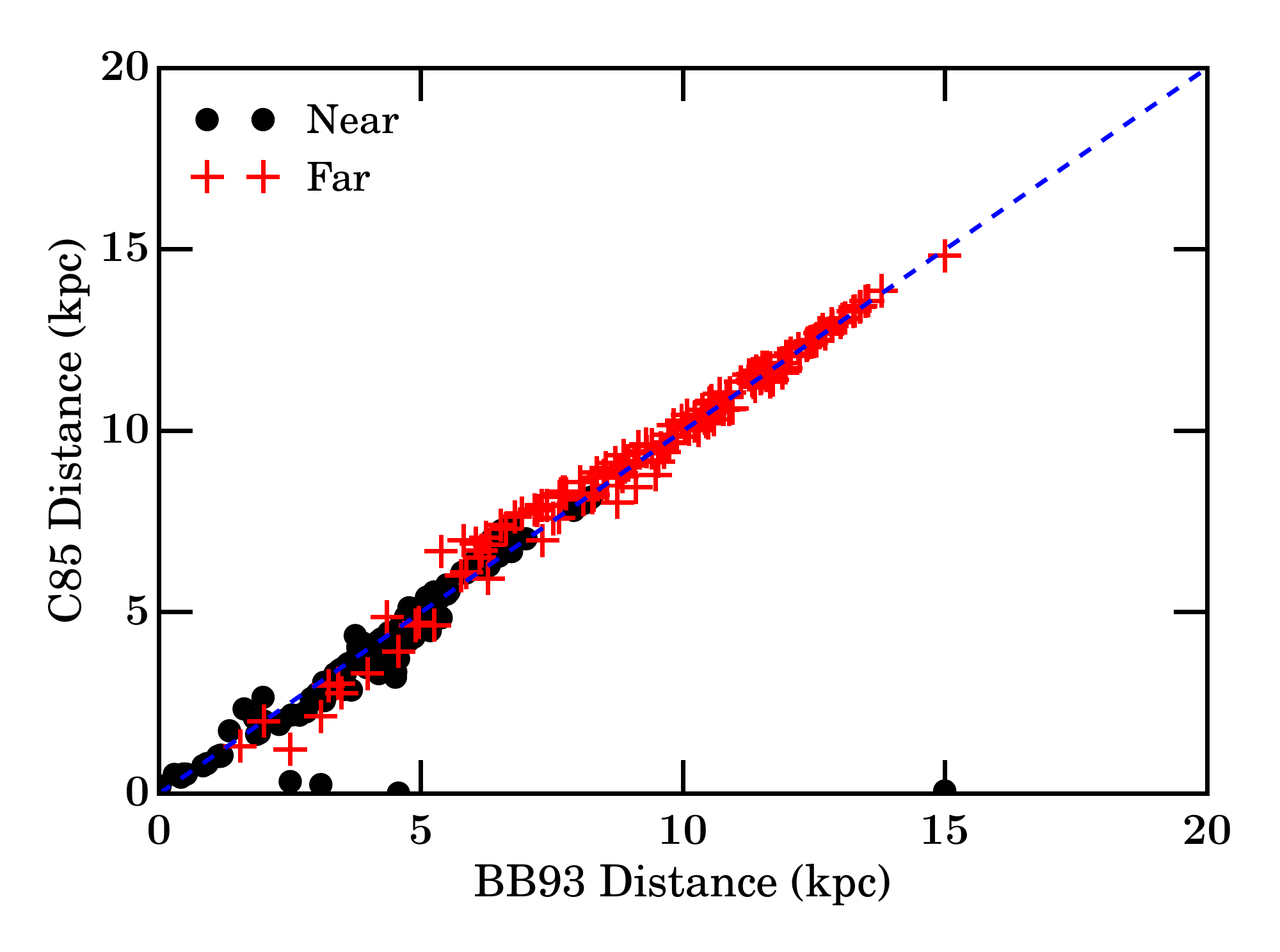}
\caption{\label{fig:dist_comp}Comparison of the kinematic distances to SFCs 
inferred from using the \citet{clemens85} and \citet{brand93} rotation curves. 
Black circles represent near 
distances, while red crosses represent far distances. The blue dashed line 
shows the relationship that would be found if the two curves gave the 
same distances. For the four outliers, where the \citet{clemens85} curve predicts 
near-zero distances, the \citet{brand93} curve predicts a unique large distance.
The \citet{clemens85} curve yields a galactocentric radius that is slightly inside the solar circle,
while the \citet{brand93} rotation curve finds a galactocentric radius that is slightly outside the solar circle.}
\end{figure}

\subsection{Rotation Curve and Kinematic Distance}
\label{ssec:rotcurve}

The rotation curve calculated by C85 is known to introduce
significant errors in the distance to the objects in the outer galaxy due to 
Perseus arm streaming motions~\citep{fich89}. 
We therefore update the \citet{lee12} distances using the rotation curve of 
\citet[][hereafter BB93]{brand93} at $R_0 = 8.5 \kpc$ and $\Theta_0 = 220 \kms$.

Figure \ref{fig:dist_comp} shows that the distances given by the two rotation curves are 
generally in good agreement. 
The four outliers with C85 `near' distances that are close to zero have C85 
solutions that are slightly inside (by $\sim 0.03\kpc$) the solar circle, 
while the BB93 solutions for these objects are slightly outside 
(by $\sim 0.12\kpc$) the solar circle. 

We reject solutions with galactocentric radius greater than 16 kpc, i.e., we are using a prior
that there is no significant massive (O star) star formation at such large radii. 
The typical error in distance is $\sim$35\%, the dominant sources of error being 
the streaming velocity of ionized gas, the expansion velocities of bubbles carved out 
by SFCs, and the motions of spiral arms~\citep{lee12}.

\subsection{Giant Molecular Clouds}
\label{ssec:gmccat}

We use the new all-sky catlaog of Galactic molecular clouds 
presented by MML16, where we describe in detail how 
clouds are identified. MML16 present a number of correlations between 
cloud properties. We provide a brief summary here.

Clouds are identified as coherent molecular structures from 
the $^{12}$CO~(J1-0) survey of \citet{dame01}. 
This survey has a modest angular resolution (7.5\,arcmin) compared to 
other observations, e.g., \citet{jackson2006}, but it has the advantage of providing 
a uniform data set covering the entire Galactic plane in longitude. 
We limit our study to $-5^\circ < b < 5^\circ$.

The identification of clouds from position-position-velocity (or [$l$, $b$, $v$]) 
cubes is challenging, as the interstellar medium (ISM) has a fractal structure,
and because unrelated clouds along a given line of sight can have similar 
projected velocities. 
Given this difficulty, it is striking that 
studies using different data sets and different structure 
identification techniques find 
relatively consistent 
scaling relations between various cloud properties,
such as size, velocity dispersion, 
mass and surface density \citep[see the review by][]{hennebelle2012a}. 
MML16 find scaling relations similar to those reported in the earlier catalogs.

MML16 employ a combination of Gaussian decomposition 
and hierarchical clustering analysis to identify clouds. 
First, every spectra of the the entire CO position-position-velocity (PPV) cube 
is decomposed into a sum of Gaussian components. 
Next, they build a cube of integrated emission $W_{\rm CO}$ 
where each grid cell ($l$, $b$, $v$) is assigned 
$W_{\rm CO}$ integrated over Gaussian components 
whose central velocities equal $v$. 
This cube is much more sparse than the original 
brightness cube as the integrated emission of each 
Gaussian component is concentrated in a single cell 
($l$, $b$, $v$) and not spread out in velocity. This 
new cube facilitates the identification of coherent 
structures down to the noise level of the data. To do 
so, they used a classical threshold descent. First, 
they identified islands in PPV space as neighboring 
cells with $W_{\rm CO}$ higher than some high threshold value. 
The threshold is progressively lowered down to 0.5 ${\rm K\,\kms}$. 
At each step of this descent, new cells are revealed. 
They are either attached to previously identified structures 
or classified as new structures. This watershed method is 
similar in spirit to {\em clumpfind} \citep{williams1994} 
or to {\em dendrogrammes} \citep{rosolowsky2008}. 
A total of 8107 coherent structures / clouds are 
identified over the whole Milky Way disk, 
recovering 91\% of the total CO emission.

As for the SFCs in this paper, the distance to each 
GMC is estimated assuming that its average 
velocity follows the Galactic rotation curve of \citet{brand93}. In the 
inner Galaxy, there is an ambiguity as two distances along the line of 
sight (dubbed near and far) have the same observed velocity. 
MML16 relied on a statistical method that 
has been used by many other studies
\citep[e.g.,][]{dame1986,solomon87,grabelsky88,garca2014}. 
The original idea was to choose the distance that provides a cloud 
physical size, $R$, that 
matches more closely the size-linewidth 
$\sigma_v \propto R^\alpha$ relation, established 
using clouds not subject to the distance ambiguity.

The size-linewidth relation appears to vary with the cloud 
column density $\Sigma$\,\citep[see e.g.,][]{heyer09}.
To account for this dependence on $\Sigma$,
MML16 select the distance that 
best matches the relation $\sigma \propto (R\Sigma)^{0.42}$.
The cloud distances chosen in this 
manner agree well with the distances of associated SFCs, 
as we show in Section \ref{ssec:correl}.

The cloud catalog provides the position in ($l$, $b$, $v$), the size, 
the velocity dispersion, and the distance to each cloud. 
It also provides the gas mass estimated as
\begin{equation}
\label{eq:CO mass}
M_{\rm g} = W_{\rm CO}^{\rm tot} \,  X_{\rm CO} \, 2 \mu m_{\rm H} \, D^2 \tan(\delta)^2
\end{equation}
where $W_{\rm CO}^{\rm tot}$ is the total CO emission of all the 
Gaussian components associated to the cloud,
$X_{\rm CO} = 2 \times 10^{20}$\,cm$^{-2}$\,K$^{-1}$\,km$^{-1}$\,s \citep{bolatto2013},
$\mu=1.36$ takes into account the contribution fron Helium, 
$m_{\rm H}$ is the mass 
of Hydrogen, $D$ is the distance to the cloud and $\delta = 0.125^\circ$ 
is the angular size of the pixel.  
We also provide the {\it WMAP} free-free and 
{\it IRAS} 100$\mu$m flux of clouds, 
measured by integrating the emission over pixels associated with each cloud.

\subsection{Cross-Correlating Star Forming Complexes with Giant Molecular Clouds}
\label{ssec:correl}

We identify matches between SFCs and GMCs if they meet the following criteria:
\begin{equation}
\sqrt{(\delta l^2 + \delta b^2 + \delta v^2)} \leq 1
\label{eq:match}
\end{equation}
where
\begin{align}
\delta l &= (l_{GMC} - l_{SFC})/\rm{max}(\sigma_{l,GMC}, R_{SFC}/D,0.5^o) \nonumber \\
\delta b &= (b_{GMC} - b_{SFC})/\rm{max}(\sigma_{b,GMC}, R_{SFC}/D,0.5^o) \nonumber \\
\delta v &= (v_{GMC} - v_{SFC})/\rm{max}(\sigma_{v,GMC}, \sigma_{v,SFC}, 7\kms).
\label{eq:lbv}
\end{align}
By $R_{SFC}/D$ we mean the angular size of the SFC measured in degrees rather than radians; 
$\sigma_{v,{\rm GMC}}$ is the RMS velocity dispersion of a GMC while $\sigma_{v,{\rm SFC}}$ 
is the half-spread velocity of SFCs (i.e., $(v_{\rm max}-v_{\rm min})/2$ along the 
bubble walls; see \citealt{lee12} for more detail). 
Out of 280 SFCs, 256 have measured median velocities $v_{\rm SFC}$.
Since the host {\it WMAP} sources 
are $\sim 2^o$ in size, we only allow SFC-GMC matches to the clouds that are 
large enough to encompass more than 10 pixels but
smaller than $2^o$ in their longest axis; larger objects 
are likely not isolated, self-gravitating clouds.\footnote{We note that 
there are a couple of SFCs with mean radii that are close to 
or exceed 2$^o$, such as SFC Nos.~110 and 111. These SFCs are associated with 
the Cygnus region within a particularly large {\it WMAP} free-free source 
(it was originally identified as three separate free-free sources by \citet{murray10} 
but \citet{lee12} merged them after a visual inspection).} 
The selection criteria on the GMC size limits the cloud count to 5469.
In summary, we perform a cross-correlation between 256 SFCs and 5469 GMCs.

Using the criteria given by equations \ref{eq:match} and \ref{eq:lbv}, 
we find that approximately half of SFCs are matched with multiple GMCs.
Visual inspection reveals that these clouds---which are often smaller in size than SFCs---trace the 
outer rim of bubbles blown by their SFC counterpart when we overlay them on 
{\it GLIMPSE} 8$\mu$m images.
It is likely that the clouds originate from a single massive cloud 
that was disrupted by stellar winds and radiation pressure from its 
massive star clusters.
These SFC-GMC ``complexes'' are often found inside the solar circle, 
so that even though multiple objects (SFCs or GMCs) coincide 
in the $(l,b,v)$-space, they can be at vastly different distances. 
We reject any GMC that is not within $0.3\,d_{\rm SFC}$ from 
the centroid of the host SFC, where $d_{\rm SFC}$ is the 
heliocentric distance of the SFC.
For SFCs without a distance measurement, we use the distance 
of the GMC that is most closely matched in the ($l,b,v$)-space.

Lastly, we make a visual inspection for the luminous SFCs 
(those with expected $M_\star \geq 10^4 M_\odot$)
to ensure that we have made a sensible cross-correlation. 
Only $\sim$10\% of the initial SFC-GMC complexes are mismatched.
Most of the mismatches stem from conflicting distance ambiguity 
resolution between the SFC catalog and the GMC catalog.

The fact that most of the luminous SFCs are matched 
with GMCs suggests a good agreement in distance ambiguity resolution
between the SFC sample (resolution by absorption line velocities) 
and the GMC sample 
(resolution by a fit to a size-linewidth-column-density relation),
enhancing our confidence in the use of the latter technique.

We identify a known cluster from \citet{morales13} in the 
direction of SFC-GMC complexes to reassign SFC Nos. 27, 28, 93 
to near distances and 
GMC Nos. 136, 171, 388, 398, 446, 523, 625, 812, 1312,
1656, 2733, 2734, 2816, 3422 
to far distances.\footnote{SFC Nos. 27 and 28 are associated with the W31 
cluster while SFC No. 93 is associated with [BDS2003] 135~\citep{bica03}.}
Distances to SFC Nos. 36 and 252 are changed to 
far distances after correcting for a bug in a code 
used to produce the catalog in \citet{lee12}.
We also manually match SFC No. 202 with GMC No. 2420, 
SFC No. 65 with GMC No. 2071 (associated with the W40 cluster~\citep{mallick13}), 
and SFC No. 251 with GMC No. 482 (associated with RCW 120~\citep{anderson10}). 
Four SFCs (Nos. 31, 35, 245, 248) had no GMC match because their 
distances were truncated to 12 kpc (because of their likely 
association with $\sim$3\kpc ring; see \citealt{lee12}); 
we recalculated their distances. 

We present the results of our cross-correlation in Table \ref{table1}.

The final result is 191 unique SFCs matched to 389 unique GMCs,
recovering 93.5$\%$ of the total SFC free-free luminosity, 83.8$\%$ of
the total SFC free-free flux, and 9\% of total GMC gas mass.  All of
the top 24 most luminous SFCs are matched to at least one GMC.  For
the rest of the paper, we identify these 389 GMCs matched to 191 SFCs
as 191 SFC-GMC complexes.

Each complex inherits the sum of the CO fluxes and gas masses of the
matched GMCs, as well as the mass-weighted mean Galactic coordinates $l,b$,
velocity $v$, heliocentric distance $d$, and galactocentric radius $R_{\rm gal}$.  
Following MML16, we calculate the angular size and surface area of each SFC-GMC
complex by solving for eigenvalues of the moment of inertia matrix:
\begin{equation}
\label{eq:inertia}
 \phi =  \left[ \begin{array}{cc}
      \sigma_l^2 & \sigma^2_{lb}\\
      \sigma^2_{lb} & \sigma_b^2
      \end{array} \right]
\end{equation}
where
%
\begin{gather}
\begin{aligned}
    \sigma_l^2 &= \frac{\sum_i M_i \sum^4_{c=1} (l_{c,i} - l_{\rm SG})^2}{4\sum_i M_i} \\
    \sigma_b^2 &= \frac{\sum_i M_i \sum^4_{c=1} (b_{c,i} - b_{\rm SG})^2}{4\sum_i M_i} \\
    \sigma^2_{lb} &= \frac{\sum_i M_i \sum^4_{c=1} (l_{c,i} - l_{\rm SG}) (b_{c,i}-b_{\rm SG})}{4\sum_i M_i}.
\end{aligned}
\end{gather}
%
Here, $i$ denotes each constituent MML16 cloud, $M_i$ is the individual 
cloud mass, and $(l_{\rm SG}, b_{\rm SG})$ are the mass-weighted 
mean Galactic coordinates of the host SFC-GMC complex.
The quantities $(l_{c,i}, b_{c,i})$ are
the Galactic coordinates of the 2 semi-major and 2 semi-minor vertices 
($c = 1$--4) of each cloud, whose semi-major axis we define as 
3$\sigma_l$ and semi-minor axis as 3$\sigma_b$.

Using the two eigenvalues $R_{\rm max}$ and $R_{\rm min}$ of $\phi$, 
we define the angular radius of each SFC-GMC complex as 
$R_{\rm ang} \equiv (R_{\rm max} R_{\rm min}^2)^{1/3}$.
We define the velocity dispersion of the SFC-GMC complex as
\begin{equation}
\label{eq:sigV_sg}
\sigma^2_{\rm v, SG} = \frac{\sum_i M_i (\delta v_i^2 + \delta_+ v_i^2 + \delta_- v_i^2)}{\sum_i M_i}
\end{equation}
where $\delta v_i = v_{\rm cent,i} - v_{\rm SG}$, 
$\delta_+ v_i = v_{\rm cent,i} + \sigma_{v,i} - v_{\rm SG}$, 
and $\delta_- v_i = v_{\rm cent,i} - \sigma_{v,i} - v_{\rm SG}$ 
with $v_{\rm cent,i}$ the central velocity 
of each cloud $i$, $v_{\rm SG}$ the mass-weighted velocity 
of the host SFC-GMC complex, and $\sigma_{v,i}$ the velocity dispersion 
of each cloud $i$. 
We have verified that the SFC-GMC complexes follow 
the $\sigma_v$-$M_g$ and $\alpha_{\rm vir}$-$M_g$ 
(where $\alpha_{\rm vir} \equiv 5 \sigma_v^2 R_g / G M_g$ 
is the cloud virial parameter in which $R_g = d \tan(R_{\rm ang})$)
relations reported in MML16.
The properties of SFC-GMC complexes are presented 
in Table \ref{table2}.

\begin{center}
\begin{deluxetable*}{lccccccccccccccc}
\tablecolumns{16}
\tabletypesize{\footnotesize}
\tablecaption{MML16-SFC match (sorted by SFC luminosities)\label{table1}}
\tablehead{
\colhead{GMC} &
\colhead{$l$} &
\colhead{$\sigma_l$} &
\colhead{$b$} &
\colhead{$\sigma_b$} &
\colhead{$v$} &
\colhead{$\sigma_v$} &
\colhead{$d$} &
\colhead{SFC} &
\colhead{$l$} &
\colhead{$b$} &
\colhead{$R$} &
\colhead{$v$} &
\colhead{$\sigma_v$} &
\colhead{$d$} \\
\colhead{No.} &
\colhead{(deg)} &
\colhead{(deg)} &
\colhead{(deg)} &
\colhead{(deg)} &
\colhead{($\kms$)} &
\colhead{($\kms$)} &
\colhead{($\kpc$)} &
\colhead{No.} &
\colhead{(deg)} &
\colhead{(deg)} &
\colhead{(deg)} &
\colhead{($\kms$)} &
\colhead{($\kms$)} &
\colhead{($\kpc$)}} 
\startdata
1726 & 336.57 & 0.14 & -0.24 & 0.14 &  -88.23 &  5.64 & 10.25 & 227 & 336.41 & -0.02 & 0.50 &  -79.00 & 20.33 & 10.63\\
1761 & 336.17 & 0.14 & 0.05 & 0.18 &  -68.67 &  7.31 & 11.07 & 227 & 336.41 & -0.02 & 0.50 &  -79.00 & 20.33 & 10.63\\
26 & 337.91 & 0.18 & 0.10 & 0.45 &  -59.09 &  8.77 & 11.62 & 228 & 337.85 & -0.20 & 0.29 &  -50.35 & 11.46 & 12.05\\
279 & 337.48 & 0.19 & -0.07 & 0.39 &  -54.86 &  6.90 & 11.81 & 228 & 337.85 & -0.20 & 0.29 &  -50.35 & 11.46 & 12.05\\
440 & 338.15 & 0.22 & -0.12 & 0.25 &  -49.99 &  7.57 & 12.09 & 228 & 337.85 & -0.20 & 0.29 &  -50.35 & 11.46 & 12.05\\
171 &  29.94 & 0.19 & -0.24 & 0.20 &  101.81 &  4.41 &  8.37 & 68 &  30.04 & -0.24 & 0.33 &   99.40 &  5.82 &  8.55\\
388 &  30.01 & 0.13 & -0.21 & 0.29 &   94.32 &  4.76 &  8.93 & 68 &  30.04 & -0.24 & 0.33 &   99.40 &  5.82 &  8.55\\
583 &  30.26 & 0.23 & -0.44 & 0.15 &  102.03 &  5.21 &  6.44 & 68 &  30.04 & -0.24 & 0.33 &   99.40 &  5.82 &  8.55\\
1054 & 305.62 & 0.32 & -0.24 & 0.27 &  -42.13 &  4.43 &  4.48 & 182 & 305.66 & -0.07 & 0.61 &  -39.10 &  9.38 &  6.24\\
678 & 358.63 & 0.24 & -0.30 & 0.09 &   -1.41 &  9.37 & 15.24 & 274 & 358.54 & -0.48 & 0.15 &   -2.60 & 11.65 & 15.24
\enddata
\tablecomments{Coordinates and velocities of SFCs and MML16 GMCs matched to each other.
The left 8 columns are properties of MML16 clouds while the right 7 columns
are properties of the matched SFCs.
 This table is published in its entirety in the electronic edition.}
\end{deluxetable*}
\end{center}

\begin{center}
\begin{deluxetable*}{lccccccccccc}
\tablecolumns{12}
\tabletypesize{\footnotesize}
\tablecaption{SFC-GMC complexes (sorted by luminosities)\label{table2}}
\tablehead{
\colhead{SFC} &
\colhead{$l$} &
\colhead{$b$} &
\colhead{$v$} &
\colhead{$\sigma_v$} &
\colhead{$R_{\rm ang}$} &
\colhead{$R_{\rm max}$} &
\colhead{$R_{\rm min}$} &
\colhead{$R_{\rm gal}$} &
\colhead{$d$} &
\colhead{$W_{\rm CO}$} &
\colhead{$M_g$} \\
\colhead{No.} &
\colhead{(deg)} &
\colhead{(deg)} &
\colhead{($\kms$)} &
\colhead{($\kms$)} &
\colhead{(deg)} &
\colhead{(deg)} &
\colhead{(deg)} &
\colhead{($\kpc$)} &
\colhead{($\kpc$)} &
\colhead{(${\rm K\,\kms}$)} &
\colhead{($M_\odot$)}} 
\startdata
227 & 336.30 & -0.05 &  -75.34 & 10.80 & 0.35 & 0.41 & 0.33 &  4.56 & 10.79 & 3.94e+02 & 9.47e+05 \\
228 & 337.90 & -0.02 &  -55.00 &  7.65 & 0.58 & 0.81 & 0.49 &  5.08 & 11.82 & 2.93e+03 & 8.49e+06 \\
68 &  30.01 & -0.25 &   98.41 &  5.34 & 0.42 & 0.52 & 0.38 &  4.44 &  8.42 & 2.08e+03 & 2.97e+06 \\
111 &  79.18 & 0.45 &   -3.21 &  3.48 & 0.95 & 1.00 & 0.93 &  8.77 &  4.25 & 9.64e+03 & 3.51e+06 \\
274 & 358.63 & -0.30 &   -1.41 &  9.37 & 0.24 & 0.47 & 0.17 &  6.75 & 15.24 & 4.83e+02 & 2.33e+06 \\
2 &   0.14 & -0.64 &   15.44 &  3.30 & 0.29 & 0.44 & 0.24 &  0.25 &  8.25 & 5.44e+02 & 7.69e+05 \\
249 & 347.78 & 0.08 &  -95.96 &  6.88 & 0.33 & 0.42 & 0.30 &  2.70 & 10.32 & 4.23e+02 & 9.33e+05 \\
110 &  76.56 & 0.25 &   -1.04 &  3.75 & 0.91 & 1.07 & 0.84 &  8.68 &  4.59 & 7.99e+03 & 3.43e+06 \\
72 &  31.02 & -0.09 &  102.25 &  5.59 & 0.45 & 0.57 & 0.40 &  4.42 &  6.69 & 1.34e+03 & 1.24e+06 \\
191 & 311.67 & 0.10 &  -50.97 &  5.53 & 0.47 & 0.56 & 0.43 &  6.52 &  7.08 & 2.60e+03 & 2.69e+06
\enddata
\tablecomments{The order of SFC Nos. appear shuffled compared to Table 1 because we adopt mass-weighted distances of 
matched GMCs for each SFC here.
This table is published in its entirety in the electronic edition.} 
\end{deluxetable*}
\end{center}

\section{Calculating Stellar Mass from Free-Free Flux}
\label{sec:flux}

We aim to measure the spread in the distribution of clouds' star 
formation efficiencies $\epsilon$ (equation \ref{eqn:efficiency})
and star formation rates $\epsilon_{\rm ff}$ (equation \ref{eqn:sfe_ff}).
We will compare the observed scatter with what is 
expected from turbulence-regulated star formation \citep[][KM05 from hereon]{krumholz05}.

The mass of stars associated with each cloud is evaluated from 
the {\it WMAP} free-free fluxes.
We provide a brief summary of photometry here; readers interested in 
more details are referred to \citet{lee12}.
For a given {\it WMAP} free-free source 
(the large $\sim$1--2$^\circ$ wide regions identified by 
their peak free-free flux in {\it WMAP}; see \citealt{murray10} 
for more detail), we first perform aperture photometry to 
compute the total free-free flux. 
The total flux is divided into constituent 
SFCs (and by extension constituent SFC-GMC complexes), 
proportional to the relative SFC 8$\mu$m fluxes computed 
from {\it Spitzer} GLIMPSE and MSX images.

The gas mass of the cloud is calculated from the CO flux $W_{\rm CO}$.
A measure of $\epsilon$ can then be probed by the flux ratio
free-free $f_\nu^{\rm br}$ over $W_{\rm CO}$:
\begin{equation}
\label{eqn:epsilon br}
\epsilon_{\rm br}\equiv {a\,f_\nu^{\rm br}/ W_{\rm CO}\over 1 + a\,f_\nu^{\rm br}/ W_{\rm CO}}
\end{equation}
where the subscript br stands for Bremsstrahlung.

To understand the constant $a$, we review 
how $f_{\nu}$'s are 
converted to the mass in live stars $M_\star$ 
and how $f_{\rm CO}$'s are 
converted to the gas mass $M_g$.

A cloud with $f_\nu^{\rm br}$ located at a distance $D$ has a luminosity 
$L_\nu^{\rm br} = 4\pi D^2 f_\nu^{\rm br}$.
Powering this luminosity requires a streaming rate of ionizing photons of 
${\cal Q} = 1.34\times 10^{26}\,(L_{\nu}/\erg\,\s^{-1}\,{\rm Hz}^{-1})~\rm{s}^{-1}$. 
The ionizing luminosity is converted to $M_\star$ using 
the ratio of ${\cal Q}$ to $M_\star$ averaged over 
the modified Muench initial mass function (IMF) from \citet{murray10}:
\be
{\left<m_{*}\right>\over \left<q\right>} = 1.6\times10^{-47} \rm{s}^{-1} \rm{M_{\odot}}.
\label{eqn:mq}
\ee
The live stellar mass in the cloud is then 
$M_\star = 1.37{\cal Q}(\left<m_\star\right>/\left<q\right>)$ 
where the numerical factor 1.37 accounts for the absorption of ionizing photons by dust 
(which compete with hydrogen atoms as a sink of ionizing photons) 
following \citet{mckee_williams97}.

The conversion between $W_{\rm CO}$ and the gas mass $M_g$ is given 
by equation (\ref{eq:CO mass}).
We can now define the constants $a$ and $b$: 
\begin{align}
\label{eqn:conv_a}
a &= \frac{M_\star/f_\nu^{\rm br}}{M_g/W_{\rm CO}} \nonumber \\
  &= \frac{4\pi\times 10^{-23} \times ({\cal Q}/L_\nu)\times 1.37 \times (\left<m_\star\right>/\left<q\right>)}{X_{\rm CO}\,2\mu m_{\rm H}\,\tan(\delta)^2},
\end{align}
where $10^{-23}$ is the conversion factor from jansky to cgs units, 
$X_{\rm CO} = 2\times 10^{20}\,{\rm cm}^{-2}\,{\rm K}^{-1}\,{\rm s}$, 
$\mu = 1.36$ to take into account helium, and $\delta = 0.125^\circ$ is
the pixel scale. Note the distance $D$ does not appear in the expression 
for $\epsilon$. Any error in the distance measurement will therefore not affect 
the scatter in the star formation efficiency.

\begin{figure}
\centering
\includegraphics[width=0.5\textwidth]{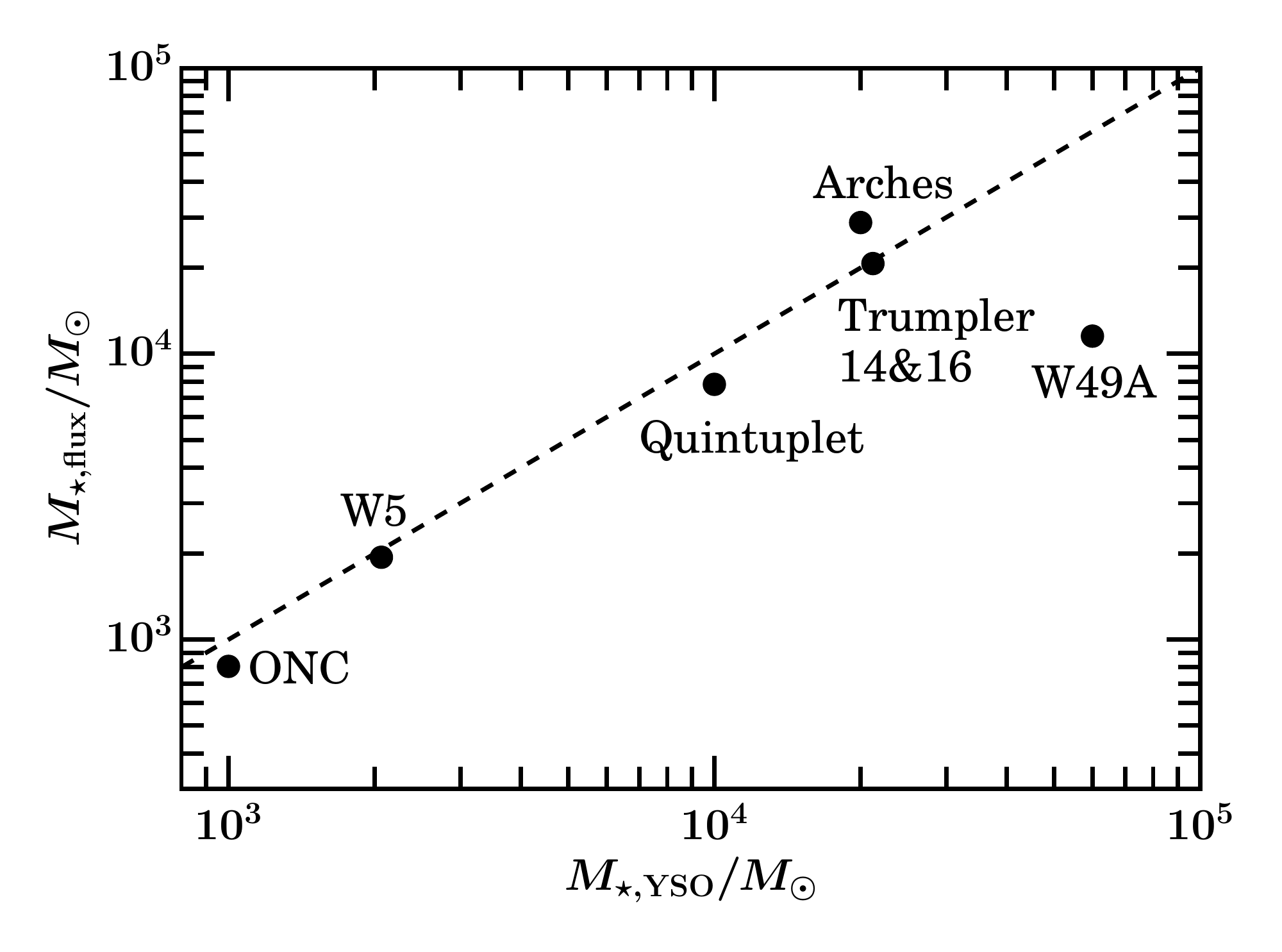}
\caption{Comparison between the stellar mass of known 
young clusters as probed by counting young stellar objects (YSOs) 
and the stellar mass in the corresponding SFCs 
as probed by free-free flux. The dashed line delineates $M_{\star,\rm flux} = M_{\star,\rm YSO}$.
We adopt YSO counts for the Orion nebula cluster (ONC) from 
\citet{DaRio12}, W5 from \citet{koenig08}, Quintuplet and Arches 
from \citet{PZ10} and references therein, Trumpler 
14 \& 16 from \citet{PZ10} and \citet{wolk11}, 
respectively, and W49A from \citet{homeier05}.}
\label{fig2}
\end{figure}

Assuming all star clusters to follow a universal IMF, our 
computed stellar mass is reliable for massive 
clusters ($M_\star \geq 10^4 M_\odot$; \citealt{krumholz15}) 
that sample their IMFs well.
There may also be variations in the IMF across 
different clusters \citep[see, e.g.,][]{dib14}.
Both the poor sampling of and the variations 
in the IMF introduce 
a scatter in the inferred stellar mass 
and by extension star formation efficiencies.
We quantify the scatter by computing 
$\sigma$ in the ratio between the stellar mass 
reported by surveys of young stellar objects (YSOs) 
and the stellar mass we compute from the free-free 
fluxes for known clusters (see Figure \ref{fig2}). 
We find $\sigma=0.22$ dex.

\begin{figure*}
\centering
\includegraphics[width=\textwidth]{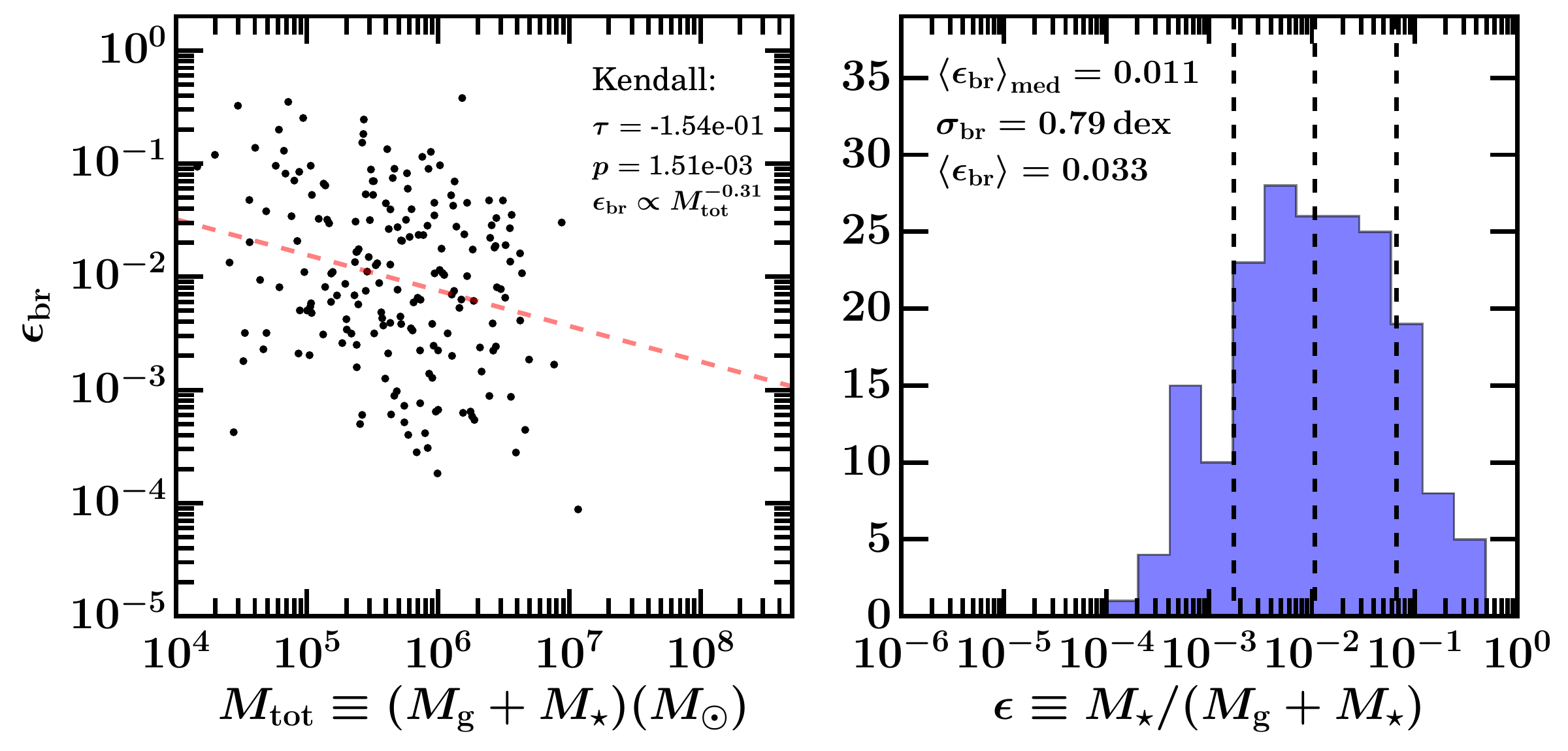}
\caption{Left: the star formation efficiency $\epsilon_{\rm br}$ 
(the subscript br stands for Bremsstrahlung)
of 191 SFC-GMC complexes plotted as a function of 
total (gas plus stellar) mass.
The correlation between $\epsilon$ and $M_{\rm tot}$ 
is statistically significant; 
we draw with the red dashed line the least-square fit 
correlation.
Right: histograms of $\epsilon$ calculated with 
free-free flux. The middle dashed line 
represents the median $\epsilon$ while 
the dashed lines to the left and right illustrate $\sigma$. 
The mean $\epsilon\sim 0.03$ is in rough agreement with that 
of nearby clouds \citep[e.g.,][]{evans09,gutermuth11}.
The full width of the distribution spans $\sim$3 orders of magnitude.
Since the p-values of Kendall's $\tau$ test is smaller than 0.05, 
we remove the annotated least-square fit between $\epsilon$ and $M_{\rm tot}$ 
before estimating the $\sigma$.}
\label{fig5}
\end{figure*}

\section{Star Formation Efficiency}
\label{sec:sfe}
 
We use the measured scatter $\sigma_{\log\epsilon}$ as one metric to
test various models of star formation rate. Using the efficiency of
star formation appears to be particularly advantageous since the
measured value of $\epsilon$ is independent of distance, so that
errors in the distance determination to either GMCs or star clusters
do not introduce any scatter in the distribution of
$\epsilon$. However, the comparison of $\sigma_{\log\epsilon}$ to the
prediction of the constant star formation rate per free-fall time
theory do require the use of a distance estimate, somewhat reducing
the attractiveness of this particular metric for our purposes.

Figure \ref{fig5} shows the star formation efficiency $\epsilon$ as a function 
of the mass of the host GMCs (left panels) and the histogram of $\epsilon$ (right panels).
We recover the result of \citet{mooney88} that the ratio of FIR to CO flux has
a very broad distribution. Specifically, we find that $\epsilon$ 
ranges over three orders of magnitude.
After correcting for the correlation between $\epsilon$ and the total mass,
we calculate the scatter $\sigma_{\log\epsilon}$ as defined in Section \ref{sec:flux}: 
about 68.3\% of the data fall within 0.79 dex.

There are a number of sources of scatter in $\epsilon$ that stem from 
measurement uncertainties.
Ionizing photons can travel surprising long distance so if the aperture chosen to 
measure the free-free flux is too small, $\epsilon_{\rm br}$ will be underestimated.
If there are nearby ($\sim 100\pc$) sources of ionizing radiation and the aperture is 
overly large, $\epsilon_{\rm br}$ will be overestimated.
Miss-identifications---regions which are not physically associated are nevertheless 
cross-correlated---will also introduce artificial scatter into the value of $\epsilon$. 

The aperture for free-free measurements are limited by the
low resolution of the {\it WMAP} free-free map. We have chosen to assign 
free-free fluxes of SFCs
according to their relative 8$\mu$m fluxes. Varying the aperture size of SFCs, 
\citet{lee12} quote an 88\% measurement error in 8$\mu$m, which we adopt for free-free fluxes 
of all GMCs. Propagating these errors in flux measurement, each cloud has an error
of $\delta \log\epsilon = 0.38$ dex.
For the entire sample of 191 star-forming clouds, measurement errors 
contribute $\Delta\sigma_{\log\epsilon_{\delta \rm flux}}=0.03$ dex. 

The conversion factor between free-free luminosities and stellar masses 
depends on the IMF. In Section \ref{sec:flux}, we estimated the 
errors introduced by the poor sampling of the IMF and the physical
variations in the IMF as 0.22 dex.
Adding the two sources of error in quadrature, we find
\be \label{eq: match err br}
\Delta \sigma_{\log\epsilon}=0.22\,{\rm dex}.
\ee
Combining all our error estimates, we arrive at our final error estimate for the 
width of the distribution of star formation efficiency:
\be \label{eq: match br}
\sigma_{\log\epsilon_{\rm br}}=0.79\pm 0.22\,{\rm dex}
\ee 
This is the first of two main observational results in this paper: the distribution 
of star formation efficiencies $\log \epsilon$ is very broad.

\section{STAR FORMATION RATE PER FREE FALL TIME $\epsilon_{\rm ff}$ }
\label{sec:sfr_ff}

The quantity $\epsilon_{\rm ff}$ is an estimate of the star 
formation rate, normalized to the free-fall time of the 
star-forming region (GMCs in our case, see equation \ref{eqn:sfe_ff}).
We will measure the scatter in the distribution of $\epsilon_{\rm ff}$ 
and compare it against what is expected from various models of star formation 
(the main test case being turbulence-regulated star formation proposed by 
\citealt[][KM05]{krumholz05} and \citealt[][HC11]{hennebelle11}).

Using the definition of $\epsilon_{\rm br}$  
from equations \ref{eqn:epsilon br}, 
equation \ref{eqn:sfe_ff} can be rewritten as
\begin{equation}
\epsilon_{\rm ff}^{\rm br} = \epsilon_{\rm br} \frac{\tau_{\rm ff}}{\left<t_{\rm ms, q}\right>},
\label{eq:eff_br}
\end{equation}
where $\left<t_{\rm ms,q}\right>\approx 3.9\Myrs$ is the 
Q-weighted main-sequence lifetime of O stars. 

We calculate the free-fall time $\tau_{\rm ff}$ using the ellipsoidal volume found 
by assuming that the length along the projected radial direction to be equal to the 
shorter of the two axes in the plane of the sky.

Unlike the measurement of $\epsilon$, the measurement of
$\epsilon_{\rm ff}$ depends on the distance to the source, via the
free-fall time $\tau_{\rm ff}$ factor in equation (\ref{eq:eff_br}). 
Thus we expect that the scatter 
in $\epsilon$ is smaller than that of $\epsilon_{\rm ff}$, and this is what we find. 
The scatter in the GMC free-fall time is 
$\sigma_{\log\tau_{\rm ff}}\approx0.27$ dex. This is much smaller than
the scatter in $\epsilon_{\rm br}$, $\sigma_{\log\epsilon_{\rm br}}=0.79$ dex. 

Part of the scatter in $\tau_{\rm ff}$ is due to uncertainties in our 
distance measurements, but most of it is intrinsic.
More precisely, $\tau_{\rm ff} \propto f_{\rm CO}^{-1/2} D^{1/2}$.
The errors in the measurement of distances are typically $\sim$35\% \citep{lee12}. 
Propagating errors in both the flux and distance measurements, 
we find errors of $\delta\log\tau_{\rm ff}=0.22$ dex for each cloud.
Combining this with the flux measurement errors in $\epsilon$,
$\delta\log\epsilon_{\rm ff}=0.44$ dex. 
For the entire sample of 191 SFC-GMCs, the measurement errors are
$\Delta\sigma_{\log\epsilon_{\rm ff}}=0.03$ dex.
Combining the measurement error with the error introduced by 
the poor sampling and the variations in the IMF $\sigma=$0.22 dex,
\be
\sigma_{\log\epsilon_{\rm ff}^{\rm br}}= 0.91\pm 0.22\,{\rm dex}
\ee

This is our second major observational result: the distribution of 
star formation rate per free fall time $\epsilon_{\rm ff}$ is very
broad.

\begin{figure*}
	\centering
    \includegraphics[width=\textwidth]{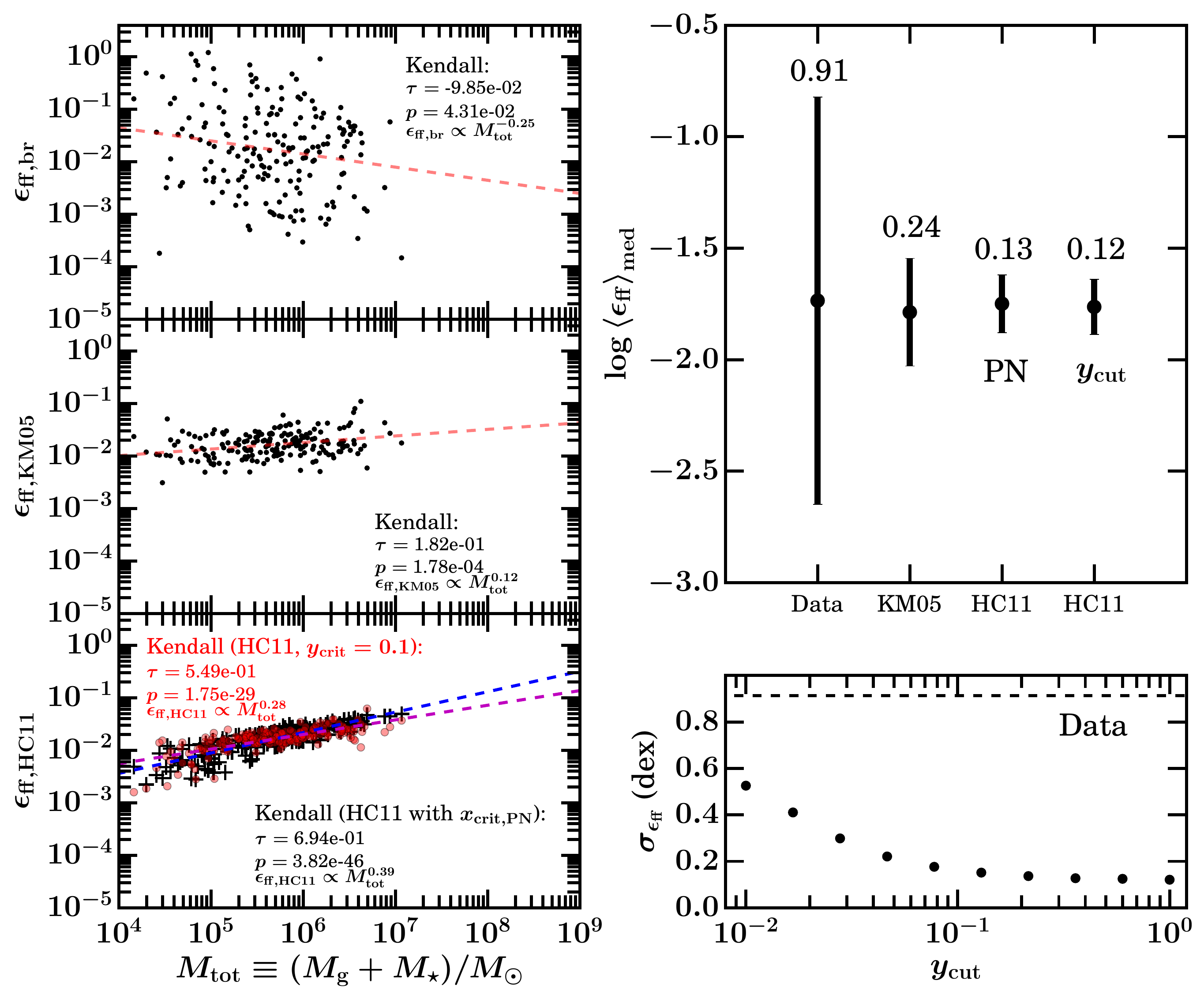}
	\caption{Top left: the star formation rate per free-fall 
    time $\epsilon_{\rm ff}$ of 191 SFC-GMC complexes 
    as a function of the total (gas + star) mass.
	We see a correlation between $\epsilon_{\rm ff}$ and 
	the total mass that is marginally statistically significant;
	the least-square fit correlation is drawn with the red dashed line.
    Middle left: $\epsilon_{\rm ff}$ predicted 
	by equation 30 of \citet{krumholz05} (equation \ref{eq:krumholz05} 
	in this paper). We see a statistically significant 
	correlation between $\epsilon_{\rm ff, KM05}$ and the total mass 
	(also drawn with the red dashed line).
    Bottom left: $\epsilon_{\rm ff}$ predicted by the 
    multi-freefall theory of star formation 
    from \citet[][labeled as HC11]{hennebelle11}, 
    their equation 8 (our equation \ref{eq:sfrffHC11_4};
    see also \citealt{federrath12}). 
    Following HC11, we compute 
    the model-inferred $\epsilon_{\rm ff,{\rm HC11}}$ using 
    two different criteria on the critical gas density over which 
    star formation occurs: the local Jeans length is equal
    to the thickness of a shocked layer 
    for a given Mach number \citep[][PN]{padoan11}; 
    and the local Jeans length is equal to 
    some prescribed fraction $y_{\rm cut}$ of the cloud size 
    \citep[][$y_{\rm cut}$]{hennebelle11}; 
    we adopt their recommended $y_{\rm cut}=0.1$.
    We see a strong positive correlation between 
    $\epsilon_{\rm ff,{\rm HC11}}$ and $M_{\rm tot}$
    in contrast to marginal negative correlation 
    observed in the data (upper left panel).
    The least-square fit correlations are drawn 
    in blue for PN and in magenta for $y_{\rm cut}$. 
    We normalized $\epsilon_{\rm ff, HC11}$ such that 
    its median matches that of the observed.
    We plot the median $\epsilon_{\rm ff}$ in the 
    upper right panel with the errorbar indicating 
    $\sigma$. The observed scatter in 
    $\epsilon_{\rm ff} = 0.91$ dex is significantly larger 
    than what any static model predicts:
    0.24 dex (KM05), 0.13 dex (HC11, PN), and 
    0.12 dex (HC11, $y_{\rm cut}$). 
    All $\sigma$'s are measured after correcting 
    $\epsilon_{\rm ff}$ for its correlation 
    with the total mass (annotated in the left panels).
    Bottom right panel shows that no $y_{\rm cut}$
    within a reasonable range can reconcile 
    HC11 model with the enormous scatter in $\epsilon_{\rm ff}$ 
    observed in the data (shown as the dashed line).
    }
    \label{fig4hc}
\end{figure*}

We can compare this directly to the prediction of KM05 and HC11, using the
properties of the GMCs we used to measure $\epsilon_{\rm ff}$
(see Section \ref{sec:cause}). We find
$\sigma_{\log\epsilon_{\rm, ff, KM05}}= 0.24$ dex 
and $\sigma_{\log\epsilon_{\rm, ff, HC11}}=0.13$ or 0.12 dex, 
depending on the choice of parameters (to be elaborated in Section \ref{sec:cause}). 
The model-inferred scatter in $\epsilon_{\rm ff}$ are 
3, 3.5, and 3.6 standard deviations from the observed value of $0.91\pm 0.22$ dex
(see Figure \ref{fig4hc}).

The fact that the measured scatter in $\epsilon_{\rm ff}$ is
significantly larger than the prediction of either KM05 or HC11 model, which
account for variations in the Mach number and virial parameter of the
host GMC (but ignores any 
{\it explicit} 
time dependence in the rate of star
formation) is a strong evidence that the model is incomplete. 
We discuss in more detail in the next section. 

We summarize the result of our calculations of $\epsilon$ and $\epsilon_{\rm ff}$ 
in Table \ref{table3}.

\begin{center}
\begin{deluxetable*}{lcccccccccc}
\tablecolumns{11}
\tabletypesize{\scriptsize}
\tablecaption{Star formation properties of SFC-GMC complexes\label{table3}}
\tablehead{
\colhead{SFC} &
\colhead{$\sigma_v$} &
\colhead{$d$} &
\colhead{$R$} &
\colhead{$Q$} &
\colhead{$M_g$} &
\colhead{$\Sigma_g$} 
&\colhead{$\epsilon_{\rm br}$} 
&\colhead{$\epsilon_{\rm ff, br}$} 
&\colhead{$\tau_{\rm ff}$} 
&\colhead{$\alpha_{\rm vir}$} \\
\colhead{No.} &
\colhead{($\kms$)} &
\colhead{($\kpc$)} &
\colhead{($\pc$)} &
\colhead{($\s^{-1}$)} &
\colhead{($M_\odot$)} &
\colhead{($M_\odot\pc^{-2}$)} &
\colhead{} &
\colhead{} &
\colhead{$(\Myr)$} &
\colhead{}} 
\startdata
227 & 10.80 & 10.79 & 66.84 & 3.64e+52 & 9.47e+05 & 6.31e+01 & 3.79e-01 & 9.14e-01 &   9.32 & 9.58e+00 \\
228 &  7.65 & 11.82 & 119.13 & 1.65e+52 & 8.49e+06 & 1.61e+02 & 3.01e-02 & 5.77e-02 &   7.41 & 9.58e-01 \\
68 &  5.34 &  8.42 & 61.64 & 9.50e+51 & 2.97e+06 & 2.23e+02 & 4.72e-02 & 5.69e-02 &   4.66 & 6.90e-01 \\
111 &  3.48 &  4.25 & 70.60 & 8.28e+51 & 3.51e+06 & 2.18e+02 & 3.52e-02 & 4.79e-02 &   5.26 & 2.83e-01 \\
274 &  9.37 & 15.24 & 64.29 & 7.24e+51 & 2.33e+06 & 1.51e+02 & 4.73e-02 & 6.86e-02 &   5.61 & 2.82e+00 \\
2 &  3.30 &  8.25 & 42.39 & 7.00e+51 & 7.69e+05 & 1.13e+02 & 1.27e-01 & 1.71e-01 &   5.22 & 6.98e-01 \\
249 &  6.88 & 10.32 & 59.93 & 6.25e+51 & 9.33e+05 & 7.36e+01 & 9.66e-02 & 1.99e-01 &   7.97 & 3.54e+00 \\
110 &  3.75 &  4.59 & 73.23 & 6.04e+51 & 3.43e+06 & 1.88e+02 & 2.68e-02 & 3.90e-02 &   5.62 & 3.49e-01 \\
72 &  5.59 &  6.69 & 52.90 & 5.81e+51 & 1.24e+06 & 8.71e+01 & 6.95e-02 & 1.03e-01 &   5.73 & 1.55e+00 \\
191 &  5.53 &  7.08 & 57.79 & 5.79e+51 & 2.69e+06 & 2.34e+02 & 3.30e-02 & 3.80e-02 &   4.45 & 7.67e-01 
\enddata
\tablecomments{The physical radius $R$ is defined as $d\tan(R_{\rm ang})$. 
The gas surface density 
$\Sigma_g \equiv M_g / (d\tan(\pi R_{\rm max} R_{\rm min}))^2$ while
the virial parameter $\alpha_{\rm vir} \equiv 5 \sigma_v^2 R / G M_g$ 
where $G$ is the gravitational constant.
This table is published in its entirety in the electronic edition.}
\end{deluxetable*}
\end{center}

\section{IMPLICATIONS OF THE LARGE SCATTER IN THE EFFICIENCY AND THE RATE OF STAR FORMATION}
\label{sec:cause}

We discuss a number of ideas regarding the physical processes that
regulate the rate of star formation in light of our observational
results.

\subsection{Turbulence regulated star formation}
\label{ssec:turbsf}

KM05 present a semi-analytic model of $\epsilon_{\rm ff}$ 
based on the idea of turbulence-regulated star formation. 
They posit that within a turbulent cloud whose density distribution 
is well-characterized by a log-normal distribution---with its width governed 
by the Mach number---only the regions of densities larger than some critical value 
will collapse to form stars. This critical value depends on both 
the virial parameter and the Mach number. Clouds 
with larger $\alpha_{\rm vir}$ and ${\cal M}$ are harder to collapse so 
the critical $\rho$ for such clouds will be larger, leading to a
smaller star formation rate per free-fall time $\epsilon_{\rm ff}$.

Assuming that the log-normal density 
distribution established by turbulence is maintained over the cloud lifetime 
(but see e.g., \citealt{vazquez-semadeni08}, \citealt{cho11}, \citealt{kritsuk11}, 
\citealt{collins12}, \citealt{federrath13}  
for the evidence of dynamic density distributions in star-forming regions),
KM05 find
\be \label{eq:krumholz05}
\epsilon_{{\rm ff},0}=0.014
\left({\alpha_{\rm vir}\over 1.3 }\right)^{-0.68}
\left({{\cal M} \over 100}\right)^{-0.32},
\ee
(their equation 30). We append a subscript 0 to $\epsilon_{\rm ff}$
to emphasize that their model {\em assumes} that the star formation rate 
is time-independent.
We see in Figure \ref{fig4hc} that 
the KM05 model expects a scatter in 
$\epsilon_{\rm ff}$, $\sigma=0.24$ dex, 
significantly smaller than the observed 
$\sigma=0.91$ dex.

\begin{figure}
    \centering
    \includegraphics[width=0.5\textwidth]{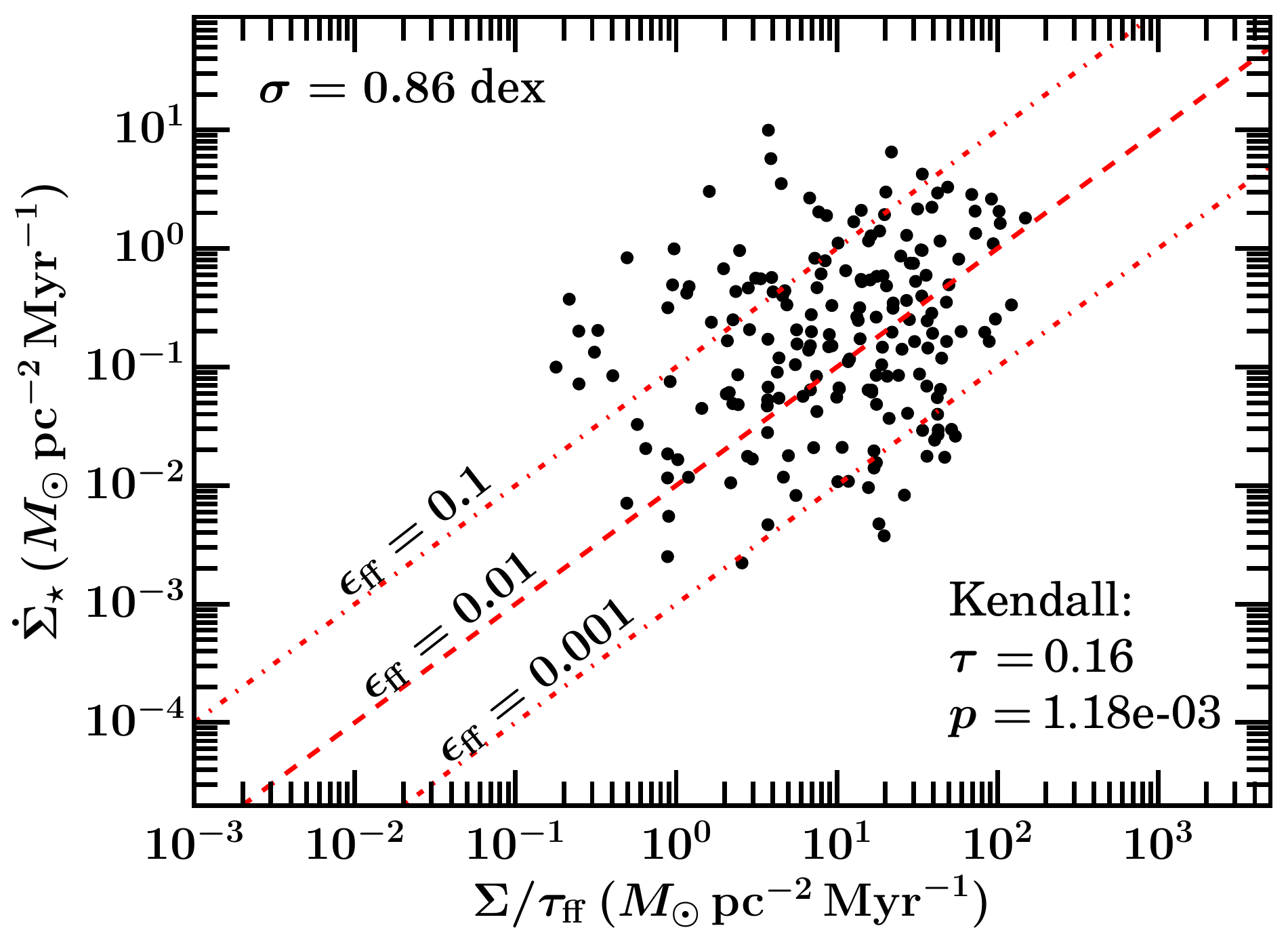}
    \caption{Star formation rate per unit area vs.~the cloud gas surface 
    density divided by the free-fall time, analogous to Figure 3 of 
    \citet{kdm12}. The red dashed line delineates the volumetric star formation law 
    proposed by \citet{kdm12} assuming a constant $\epsilon_{\rm ff}=0.01$. 
    We draw two different constant $\epsilon_{\rm ff}$ (0.1 and 0.001) 
    in red dot-dashed lines for reference.
    We find a significant correlation between 
    $\Sigma_\star$ and $\Sigma/\tau_{\rm ff}$. However, the scatter about 
    the volumetric star formation law 
    $\dot\Sigma_\star \propto \Sigma/\tau_{\rm ff}$ is large: 
    0.86 dex.}
    \label{fig8}
\end{figure}

The KM05 model has been criticized for 
characterizing an entire cloud 
and its collapsing substructure by 
a single free-fall timescale. 
Star-forming substructures have 
varying densities and therefore 
different free-fall times
(see e.g., \citealt{hennebelle11}, 
\citealt{federrath12}, and references therein).
\citet{hennebelle11} present an analytic model of $\epsilon_{\rm ff}$,
which takes into account not only the different free-fall time 
for each collapsing structure but also the 
recycling of turbulent flow over a dynamical time
\citep[see also][]{federrath12,hennebelle13}:
\begin{equation}
\epsilon_{\rm ff} \propto e^{3\sigma_{\rho}^2/8}\left[1+{\rm erf}\left(\frac{\sigma_{\rho}^2-\ln x_{\rm crit}}{2^{1/2}\sigma_{\rho}}\right)\right]
\label{eq:sfrffHC11_1}
\end{equation}
where $\sigma_{\rho}$ is the spread in the density distribution, set by 
the cloud Mach number:
\begin{equation}
\sigma_{\rho}^2 = \ln(1 + b^2{\cal M}^2)
\label{eq:sfrffHC11_2}
\end{equation}
($b = 0.25$ for purely solenoidal driving and $b=1.0$ 
for purely compressive driving; we take $b=0.5$)
and $x_{\rm crit} \equiv \rho_{\rm crit}/\rho_0$ is 
the critical density over which substructure begins to collapse
normalized by the average bulk density of the cloud $\rho_0$.
We experiment with two criteria for collapsing:
if the local Jeans length becomes comparable 
to the local shock width \citep[][PN criterion]{padoan11}:
\begin{equation}
x_{\rm crit} \simeq 0.067 \theta^{-2} \alpha_{\rm vir} {\cal M}^2
\label{eq:sfrffHC11_3}
\end{equation}
with $\theta = 0.35$ and 
if the local Jeans length becomes comparable 
to some prescribed fraction $y_{\rm cut}$ of the cloud size 
\citep[][$y_{\rm cut}$ criterion]{hennebelle11}:
\begin{equation}
x_{\rm crit} = \frac{1}{5 y_{\rm cut}} \frac{\alpha_{\rm vir}}{{\cal M}^2}(1 + {\cal M}y_{\rm cut}^p).
\label{eq:sfrffHC11_4}
\end{equation}
Here, $p$ is the power-law scale from 
the size-linewidth relationship $\sigma \propto R^p$; we 
adopt p = 0.5.
The normalization of equation \ref{eq:sfrffHC11_1} represents 
the efficiency at which a clump gas converts to a star, and 
we adjust it so that the median model-inferred 
$\epsilon_{\rm ff}$ matches that of the observed. 
The match warrants an unusually small 
normalization: 0.009 for the PN criterion and 
0.002 for the $y_{\rm cut}$ criterion, 
compared to the usual 0.02--0.05.

These multi-freefall models of star formation show stronger dependence 
of $\epsilon_{\rm ff}$ on the Mach number compared 
to the KM05 model. It has been argued 
that the observed scatter in $\epsilon_{\rm ff}$ of Milky Way clouds 
can be explained if the clouds' Mach number or the scale of 
turbulence driving vary by $\gtrsim2$ orders of magnitude
\citep[e.g.,][]{federrath13b,chabrier14}.
Figure \ref{fig4hc} demonstrates that the velocity dispersion 
or the size of the clouds simply do not vary enough to reconcile 
even the multi-freefall model 
with the observed scatter in $\epsilon_{\rm ff}$ 
of the MML16 clouds. 
No reasonable choice 
of $y_{\rm cut}$ can 
reproduce the large $\sigma=0.91$ dex.
Furthermore, HC11 models predict 
a strong positive correlation between $\epsilon_{\rm ff}$ 
and the total mass in contrast to a weak negative correlation 
seen in the data.

Three facts emerge from our analysis that 
hints at the controlling parameters 
of star formation other than $\alpha_{\rm vir}$, ${\cal M}$, 
and the scale of turbulence driving.
First, the measured scatter is significantly larger 
than any theoretically expected scatter.
Second, there is little systematic offset between the observed 
and the KM05 distribution of $\epsilon_{\rm ff}$.
Third, an unusually small gas-to-core efficiency is required
to match the median $\epsilon_{\rm ff}$ expected from the multi-freefall 
models to that observed.
All three facts suggest 
the distribution of Milky Way cloud ${\cal M}$ and size 
is far too narrow to explain the observed scatter in $\epsilon_{\rm ff}$ 
with what the multi-freefall models predict.

Both the larger scatter and the lack of systematic offset between the observed 
and theoretical distribution of $\epsilon_{\rm ff}$ 
is also evident in Figure \ref{fig8}.
The figure shows the star formation rate per unit area $\dot\Sigma_\star$ 
plotted against the ratio between the gas surface density 
and the free-fall time, $\Sigma/\tau_{\rm ff}$, a comparison advocated
by \citet{kdm12}.

Like \citet{kdm12}, we find a statistically significant correlation
between $\dot\Sigma_\star$ and $\Sigma/\tau_{\rm ff}$.  A least-square
fit produces $\dot\Sigma_\star \propto (\Sigma/\tau_{\rm
  ff})^{0.3}$, compared to the  linear relationship (the volumetric star
formation law) reported by \citet{kdm12}.

\begin{figure}
    \centering
    \includegraphics[width=0.5\textwidth]{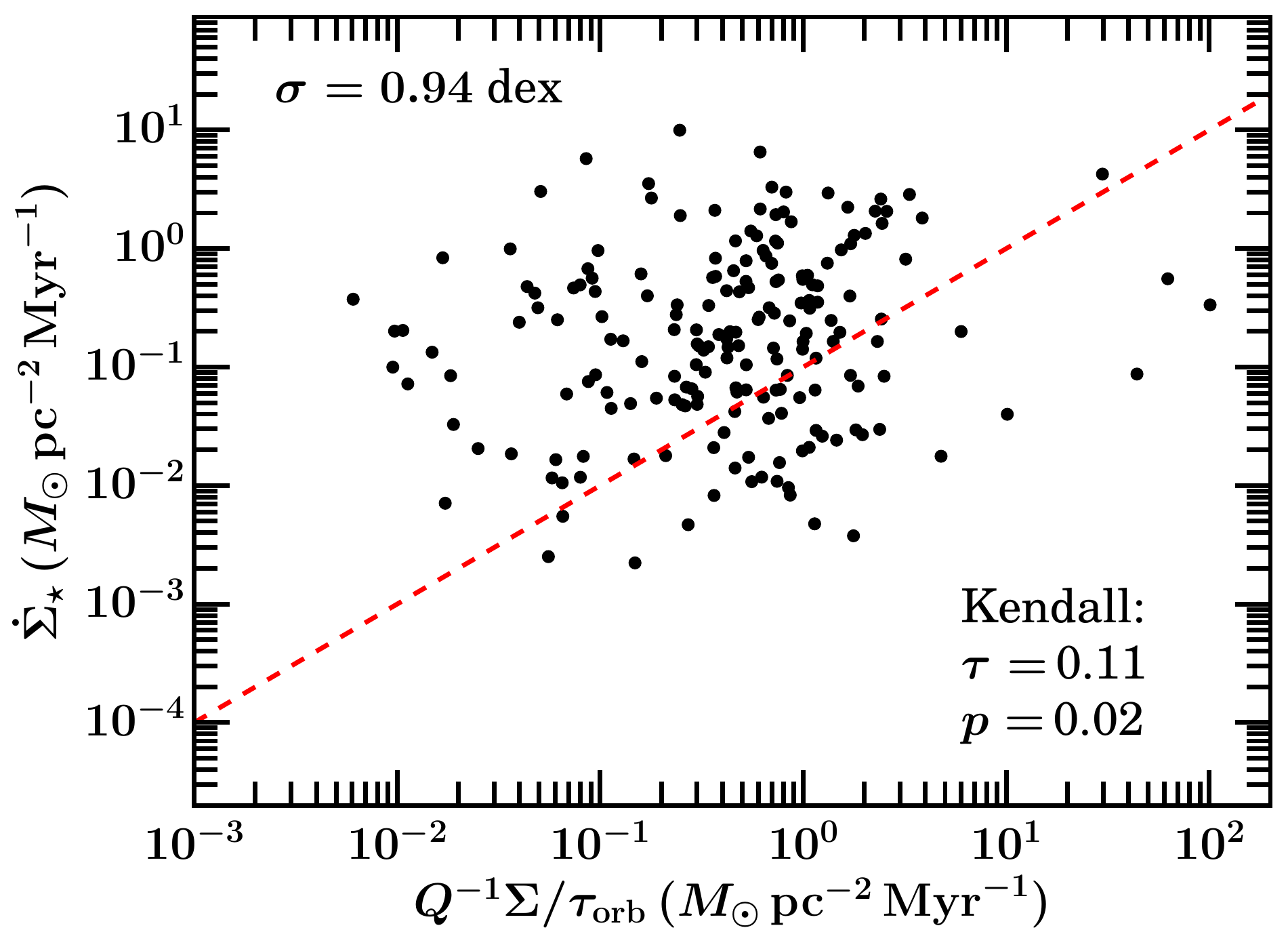}
    \caption{Star formation rate per unit area $\dot\Sigma_\star$ 
    vs.~the cloud surface density divided by the collision time 
    $\sim \tau_{\rm orb}Q$.
    The orbital time $\tau_{\rm orb} = 2\pi R_{\rm gal} / 220\kms$ 
    where $R_{\rm gal}$ is the galactocentric radius and 
    we assume a flat rotation curve $v(r) = 220\kms$.
    The red dashed line illustrates equation \ref{eq:sf_col} 
    with $\epsilon_{\rm col} = 0.1$, 
    assuming a flat rotation curve ($\beta = 0$).
    We see a strong correlation between 
    $\dot\Sigma_\star$ and $Q^{-1}\Sigma/\tau_{\rm orb}$ 
    but the scatter about the model is still large: 0.92 dex.
    }
    \label{fig9}
\end{figure}

We interpret the difference in the power-law index as the result of
the large scatter in the observed value of $\epsilon_{\rm ff}$.
As Figure \ref{fig8} attests, 
the data points span more than an order of magnitude both 
above and below $\epsilon_{\rm ff}=0.01$, well beyond the expected 
uncertainty $\sim$3 predicted by \citep{kdm12}.
We find a dispersion
about the volumetric law of 0.86 dex. 
Similarly large scatters in star
formation rates are observed in local molecular gas from Gould's belt
\citep{evans14} and in ATLASGAL clumps \citep{heyer16}. 
The sample of clouds assembled by \citet{evans14} 
falls squarely within the range $\epsilon_{\rm ff}=0.1$ and 
$\epsilon_{\rm ff}=0.001$. A few ATLASGAL clumps analyzed by \citet{heyer16} 
fall below $\epsilon_{\rm ff}=0.001$ but no 
clumps lie beyond $\epsilon_{\rm ff}=0.1$.
Our GMC sample is
more complete than the clump samples in either the Gould's belt or
ATLASGAL clumps, which may explain part of the difference in the
dispersions found in the different surveys.

\subsection{Collision-induced Star Formation}
\label{ssec:collision}

An alternate form of star formation law envisions 
cloud-cloud collision to regulate the star formation rate
on approximately orbital timescales 
\citep[see e.g.,][and references therein]{tan00, suwannajak14}:
\begin{equation}
\dot\Sigma_\star \simeq \epsilon_{\rm col} Q^{-1}(1-0.7\beta)\Sigma / \tau_{\rm orb}
\label{eq:sf_col}
\end{equation}
where $\epsilon_{\rm orb}=0.1$ is the rate at which gas is converted to 
stars for each collision, $Q \sim \sqrt{\alpha_{\rm vir}}$ the Toomre parameter, 
$\tau_{\rm orb}$ the orbital time, and 
$\beta$ is the logarithmic derivative of the velocity profile. 
Assuming a flat rotation curve ($\beta=0$), we 
show in Figure \ref{fig9}
that our star-forming clouds scatter about the 
collision-induced star formation 
law with a dispersion of $\sigma=$0.94 dex. 
As in Figure \ref{fig8}, the full min-mid or mid-max scatter 
is more than an order of magnitude; it is unlikely that 
the shear-velocity dependent term $(1-0.7\beta)$ 
vary by more than a factor of order unity.

As with the volumetric star formation law, the large scatter about the
collision-induced star formation law suggests there are extra
parameters that control the rate at which gas is converted into stars.

\begin{figure}
    \centering
    \includegraphics[width=0.5\textwidth]{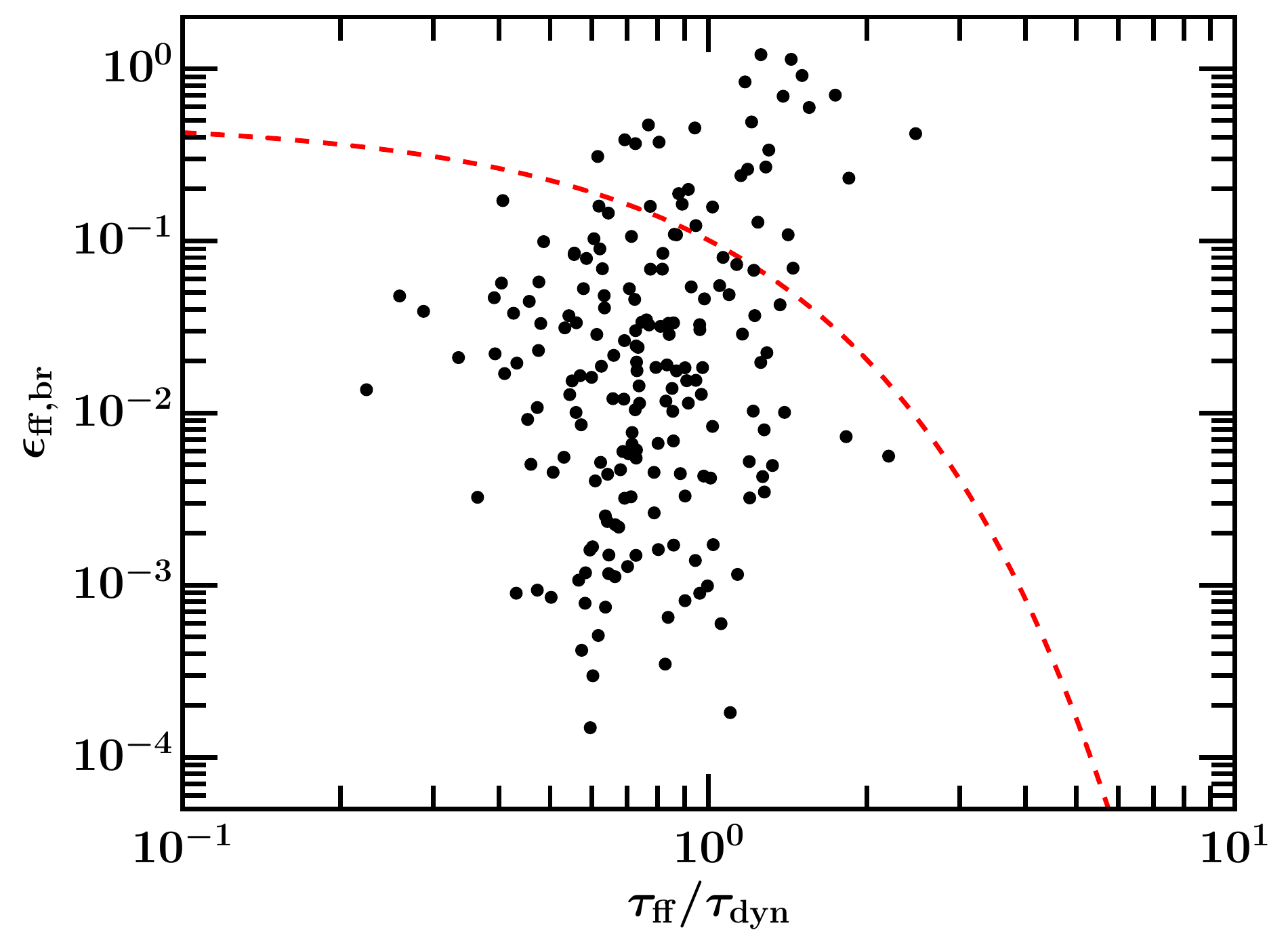}
    \caption{Star formation rate per free-fall time $\epsilon_{\rm ff}$
    plotted against the ratio between the free-fall and the dynamical time, 
    $\tau_{\rm ff}/\tau_{\rm dyn}$. The star formation law in 
    equation \ref{eq:eff_padoan} is drawn as red dashed lines. 
    The time ratio (equivalently, the square-root of the virial parameter)
    may be an important parameter related to the 
    maximum possible $\epsilon_{\rm ff}$, but by itself 
    it does not explain the large scatter in $\epsilon_{\rm ff}$.}
    \label{fig10}
\end{figure}

\subsection{SFR Parametrized by Virial Parameter}
\label{ssec:sfr-vir}

\citet{padoan12} studied magneto-hydrodynamic simulations of
turbulent, star-forming gas, finding that $\epsilon_{\rm ff}$ depends
exponentially on the ratio between the free-fall and the dynamical
time (which is equivalent to the square-root of the virial parameter):
\begin{equation}
\epsilon_{\rm ff} \simeq \epsilon_{w}e^{\left(-1.6~\tau_{\rm ff}/\tau_{\rm dyn}\right)},
\label{eq:eff_padoan}
\end{equation}
where $\epsilon_{w}$ accounts for the mass loss due to 
proto-stellar winds and outflows, 
and $\tau_{\rm dyn}$ is the dynamical time in their simulations. 

Figure \ref{fig10} shows the star formation rate per free-fall 
time for star-forming clouds, plotted
as a function of $\tau_{\rm ff}/\tau_{\rm dyn}$ 
where $\tau_{\rm dyn} = R_g / \sigma_{\rm GMC}$ is
the dynamical time of the host cloud. 
We do see a hint of a decrease in the upper envelope of the distribution of
$\epsilon_{\rm ff}$ with increasing $\tau_{\rm ff}/\tau_{\rm dyn}$. 
However, many GMCs are more efficient at producing stars than what 
equation \ref{eq:eff_padoan} would suggest. More strikingly, 
the $\epsilon_{\rm ff}$ values
of GMCs span $\sim$3 orders of magnitude {\em below} the 
prediction by \citet{padoan12}. 

\citet{padoan12} noted that $\epsilon_{\rm ff}$ 
depended on the Alfv\'enic Mach number ${\cal M}_a$, with lower Alfv\'enic Mach number 
(stronger magnetic field for a given strength of turbulence) 
yielding smaller $\epsilon_{\rm ff}$, but only for ${\cal M}_a\gtrsim 5$; 
below this value, $\epsilon_{\rm ff}$ increases again (see their Figure 3). 
Their simple fitting formula (our equation \ref{eq:eff_padoan}) 
corresponds to the minimum star formation rates in their models 
with ${\cal M}_a\approx 5$. Can the variation in the magnetic field strength 
explain the large scatter in $\epsilon_{\rm ff}$? 

Variations in ${\cal M}_a$ may explain those points that lie above
the red dashed line (corresponding to equation \ref{eq:eff_padoan}) in 
Figure \ref{fig10}, but the extreme low values seen 
below the red dashed line cannot be explained by variations
in the magnetic field strength. 
The Alfv\'enic Mach number can be estimated 
as ${\cal M}_a \sim \alpha_{\rm vir} (M_g/M_\Phi)$ 
where $\alpha_{\rm vir}$ is the virial parameter and 
$M_\Phi = 0.12\pi B (R_g/2)^2 / \sqrt{G}$ is the magnetic critical mass 
where $B$ is the magnetic field strength. SFC-GMC clouds typically have 
volumetric number densities of $n_{\rm H} \sim\,$10--200$\,\,{\rm g\,\,cm^{-3}}$ 
so their $B\sim$1--10$\mu$G \citep{crutcher12}; we find that these clouds 
have $M_g/M_\phi \sim\,$10--100.
The median $\alpha_{\rm vir}\sim$0.76 with 
a scatter of 0.32 dex. We estimate ${\cal M}_a\sim\,$4--200.

The simulations of \citet{padoan12} show that an order of magnitude change in 
${\cal M}_a$ results in a factor of $\sim$3 change in $\epsilon_{\rm ff}$.
Taking the variations in ${\cal M}_a$ of the star-forming clouds into account, 
we expect an upward scatter from the reference value 
given by equation \ref{eq:eff_padoan} by factors of $\sim$10. 

\subsection{Constant vs.~Time-dependent star formation rate per free fall time}
\label{ssec:time_dependent}

All four models described in 
Section \ref{sec:cause}---single-freefall turbulence-regulated, 
multi-freefall turbulence-regulated,
collision-induced, 
and parametrization by virial parameter---assume a
star formation rate that has no explicit dependence on time.
These models predict that any observed 
scatter in $\epsilon_{\rm ff}$ should arise from variations 
in the internal properties of clouds (e.g., virial parameter, Mach
number, Alfv\'enic Mach number, or 
free-fall time) or large scale motions.
We have demonstrated above that the distribution in the 
observed $\epsilon_{\rm ff}$ of Milky Way GMCs 
is far too broad to be explained by any of these models 
(equations \ref{eq:krumholz05}, \ref{eq:sfrffHC11_4}, \ref{eq:sf_col}, and \ref{eq:eff_padoan}).

A number of authors have suggested that the star formation rate on 
scales of GMCs and smaller increases systematically with time 
(see e.g., \citealt{palla99}, \citealt{gutermuth11} 
and \citealt{murray11} for observational evidence 
or see e.g., \citealt{lee15}, \citealt{murray15}, 
and \citealt{dmurray15} for theoretical studies). 
Sampling clouds at different evolutionary stages 
with time-varying star formation rate
may give rise to the broad distributions 
in $\epsilon$ and $\epsilon_{\rm ff}$.
In this section, we allow $\epsilon_{\rm ff}$ to be 
time-dependent and compare the expected width in the distribution of 
$\epsilon_{\rm ff}$ to that observed.
We note that the explicit time-dependence of $\epsilon_{\rm ff}$ 
arises from the time-varying turbulent 
structure (i.e., the size-linewidth relation itself 
changes with time due to the interplay between 
turbulence and gravity; 
see \citealt{goldbaum11} and \citealt{murray15}).
This explicit time-dependence should not be confused 
with the implicit time-dependence portrayed by \citet{hennebelle11} 
whose model accounts for the re-assembly of turbulent structure 
that is {\it static} (i.e., the turbulent flows follow the same 
size-linewidth relation each time they are recycled).

We proceed to test the effect of explicit 
time evolution of $\epsilon_{\rm ff}$ on the distribution of $\epsilon_{\rm ff}$. 
First, we generalize the \citet{krumholz05} model to allow 
for time-variable $\epsilon_{\rm ff}$,
\be
{dM_*\over dt}=\epsilon_{{\rm ff},0}\left({t\over\tau_{\rm ff}}\right)^\delta {M_g\over \tau_{\rm ff}},
\ee
where $t=0$ corresponds to the time at which the first star forms, and 
$\delta=0$ corresponds to the constant $\epsilon_{\rm ff}$ model.

Next, we use this prescription for $\epsilon_{\rm ff}$ 
in the model of \citet{feldmann11} 
which describes the concomitant evolution of the 
stellar mass and the gas mass of a star-forming cloud:
\begin{eqnarray}\label{eq:feldmann}
{dM_g\over dt}&=&-\epsilon_{{\rm ff},0}\left({t\over \tau_{\rm ff}}\right)^\delta 
{M_g\over\tau_{\rm ff}}-\alpha M_*+\gamma,
\\
{dM_*\over dt}&=& \epsilon_{{\rm ff},0}\label{eq:feldmann2}
\left({t\over \tau_{\rm ff}}\right)^\delta 
{M_g\over\tau_{\rm ff}}.
\end{eqnarray}
The quantity $\alpha$ parametrizes the rate at which stellar 
feedback disrupts the host GMC, 
while $\gamma$ is the (possibly time dependent) rate of 
gas accretion onto the GMC.

\begin{figure*}
\centering
\includegraphics[width=\textwidth]{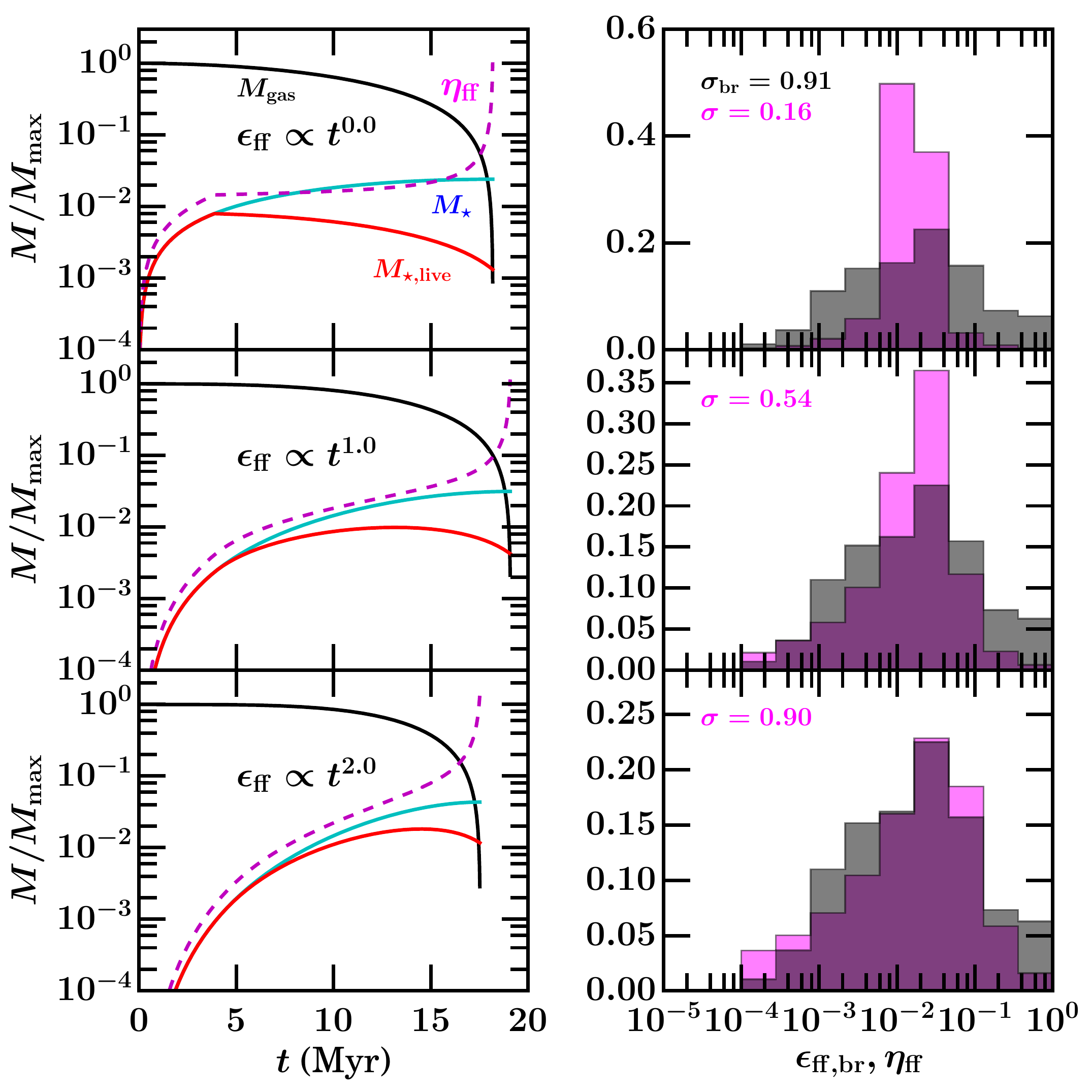}
\caption{Left: the evolution of the GMC gas mass 
$M_g$ (black line), stellar mass $M_\star$ (cyan line), 
and live stellar mass $M_{\star,\rm live}$ (red line; those with effective 
lifetime of $\left<t_{\rm ms,q}\right>\sim 4\Myrs$), 
together with the star formation rate per free-fall time
$\eta_{\rm ff} \equiv [M_{\star,\rm live} / (M_g + M_{\star,\rm live})] (\tau_{\rm ff} / \left<t_{\rm ms,q}\right>)$ 
(magenta dashed line) for three different 
values of $\delta =$ 0, 1, 2 in the model described by equations 
\ref{eq:feldmann} and \ref{eq:feldmann2}.
We use the observed median free-fall time $\tau_{\rm ff} \simeq 6.7\Myrs$.
Right: the predicted distributions of 
$\eta_{\rm ff}$ (magenta histograms) together with
the observed distributions (gray histograms). 
We assume no gas accretion (results for constant 
accretion models are similar). 
A quadratic growth $\epsilon_{\rm ff} \propto t^2$ (bottom panels) reproduces 
the observed scatter in $\epsilon_{\rm ff}$ the best.}
\label{fig11}
\end{figure*}

Equations \ref{eq:feldmann} and \ref{eq:feldmann2} 
are coupled ordinary differential equations (ODEs)
whose solutions are $M_g(t)$ and $M_\star(t)$. 
Massive stars that contribute most to the ionizing 
radiation typically live for $\left<t_{\rm ms,q}\right>\simeq 3.9\Myrs$ 
so we define the mass in these ``live'' stars as 
$M_{\rm \star, live} (t) = M_\star(t) - M_\star(t - \left<t_{\rm ms,q}\right>)(t)$.
We can then write the star formation efficiency as 
$\epsilon(t) \equiv M_{\rm \star, live}(t) / (M_g(t) + M_{\rm \star, live}(t))$ 
and the star formation rate per free fall time as 
$\eta_{\rm ff}(t) \equiv \epsilon(t) \tau_{\rm ff} / \left<t_{\rm ms,q}\right>$.
Note that $\eta_{\rm ff}$ is analogous to the 
observed $\epsilon_{\rm ff}$, not to be confused with 
$\epsilon_{\rm ff,0} (t / \tau_{\rm ff})^\delta$.

We illustrate the predicted evolution and the distribution of
$\eta_{\rm ff}$ of this model in Figure \ref{fig11} for three different
$\delta=0,1,2$ assuming no accretion ($\gamma=0$).  
We set $\tau_{\rm ff}$ to the median GMC free-fall 
time $\tau_{\rm ff} \sim 6.7\Myrs$
and set $\alpha=3.5$ so that the cloud disperses in $\sim 20\Myrs$
e.g., \citealt{williams97} and \citealt{murray11}. 
In Section \ref{ssec:lifetime} below we obtain a mean 
cloud lifetime using MML16 clouds of 21--24$\Myrs$. 
To isolate the effect of $\delta$, we assume all
the host GMCs have the same $\alpha_{\rm vir}$ and ${\cal M}$, or more
precisely, the same intrinsic value of $\epsilon_{\rm ff,0}$. We
account for variations in these parameters after integrating the ODEs
by adding the variations in quadrature.

The distribution of $\eta_{\rm ff}$ that we calculate from the model with
$\delta=0$ (the constant $\epsilon_{\rm ff}$ model) features a very
sharp peak.  Because the lifetime of GMCs ($\sim20\Myrs$) is
substantially longer than the lifetime of live stars ($\sim4\Myrs$),
the distribution in $\eta_{\rm ff}$ is dominated by the period in which
both the live stellar mass and the host GMC mass are roughly constant
with $\epsilon \sim \epsilon_{\rm ff, 0} (4 \Myrs / \tau_{\rm ff})$ 
and therefore a constant $\eta_{\rm ff} \sim \epsilon_{\rm ff, 0}$.
The small scatter (0.16 dex) arises partly from the initial rise in
the live stellar mass, and partly from the final rapid decline in GMC
mass around $\sim20\Myrs$, when the stellar feedback disrupts the host
GMC. 

To make a fair comparison to the observed scatter in 
$\epsilon_{\rm ff}$, we add in quadrature the scatter
from the initial and final transients, 0.16 dex, and the dispersion in
$\epsilon_{\rm ff,0}=0.24$ dex, to find a total predicted dispersion
of $\sigma_{\log\eta_{\rm ff}}=0.29$.
Compared to the observed $\sigma_{\rm br} = 0.91 \pm 0.22$ dex, 
the scatter predicted by the models of 
time-independent $\epsilon_{\rm ff}$ 
is too small by 2.8 standard deviations.

Allowing for a monotonic increase in $\epsilon_{\rm ff}$ with time significantly 
broadens the distribution in $\epsilon$ and $\epsilon_{\rm ff}$ 
(see second and third rows in Figure \ref{fig11}). 
Not only does the mass in live stars monotonically increase over most of the cloud 
lifetime, it also increases more gradually from time zero.
When $\delta=1$ (as predicted by \citealt{lee15} and \citealt{murray15}), 
we find $\sigma=0.54$. Adding this in quadrature to the scatter in 
GMC properties, we find a total dispersion of 
$\sigma_{\log\eta_{\rm ff}}=0.59$
(1.4-sigma away from $\sigma_{\rm br}$).
Similar calculations show a total dispersion of 
$\sigma_{\log\eta_{\rm ff}}=0.93$ for $\delta=2.0$ 
(within one standard deviation from $\sigma_{\rm br}$).

Models assuming constant star formation rate per free-fall time 
predict too narrow a distribution in $\epsilon_{\rm ff}$.
Using the simple model of \citet{feldmann11}, 
$\epsilon_{\rm ff} \propto t^2$---equivalently, the stellar mass 
increases as $M_*(t)\sim t^3$---fits the data better.
Our result demonstrates that a time-varying 
star formation rate is in better agreement 
with the data compared to time-independent models. 
Determining the exact form 
of $\epsilon_{\rm ff}$ that best fits the observation
will require a careful parameter study which 
is beyond the scope of this paper.

We note that all predicted distributions of 
$\eta_{\rm ff}$ 
(whether the star formation rate is assumed to be time-dependent or constant) 
feature a deficit towards the high end and an excess towards the low end
compared to the observations.
The simple model of 
\citet{feldmann11} assumes that feedback from stars 
begins to destroy the host GMC as soon as the stars form. If the feedback
takes the form of gas pressure in HII regions, or of radiation pressure, the host GMC
will not be significantly affected until either pressure (or their sum) overcomes the 
pressure associated with the self-gravity of the GMC. Altering the model to account for
this threshold effect will extend the distribution of $\epsilon$ to lower values. We leave
this and similar model building efforts to future work.

\subsection{Cloud Lifetimes}
\label{ssec:lifetime}

We estimate the average cloud lifetime using the star-forming clouds 
(191 SFC-GMC complexes) and non-star-forming clouds that are 
massive enough---and gravitationally bound---to birth stars.
We show that clouds indeed live substantially longer than $\sim4\Myrs$. This 
separation of time scales is crucial if we are to use the width of the
distribution of $\epsilon$ (or $\epsilon_{\rm ff}$) to distinguish between constant 
star formation rate models and models which allow for variations of 
the star formation rate with time.

Clouds that harbor at least one SFC are likely near the end of their lives:
their SFCs have already carved out bubbles of size 10--100 pc. 
The lifetime of clouds can then be estimated by multiplying 
the effective lifetime of live stars by the ratio of the 
total number of potentially star-forming clouds to the 
number of clouds that have SFCs. 
We define potentially star-forming clouds as those 
that are both massive enough to birth 
Orion Nebula Cluster (i.e., $M_{\rm gas} \geq M_{*} \sim 1000 M_\odot$) 
and gravitationally bound ($\alpha_{\rm vir} < 3.3$; see Appendix of MML16); 
there are 1014 such clouds.
The average cloud lifetime is then $((1014+191)/191) \times 4 \Myrs\sim 24\Myrs$. 
The lifetime reduces to 21$\Myrs$ if we place the cloud lower mass limit 
at $10^4 M_\odot$ instead.

\begin{figure}
\centering
\includegraphics[width=0.5\textwidth]{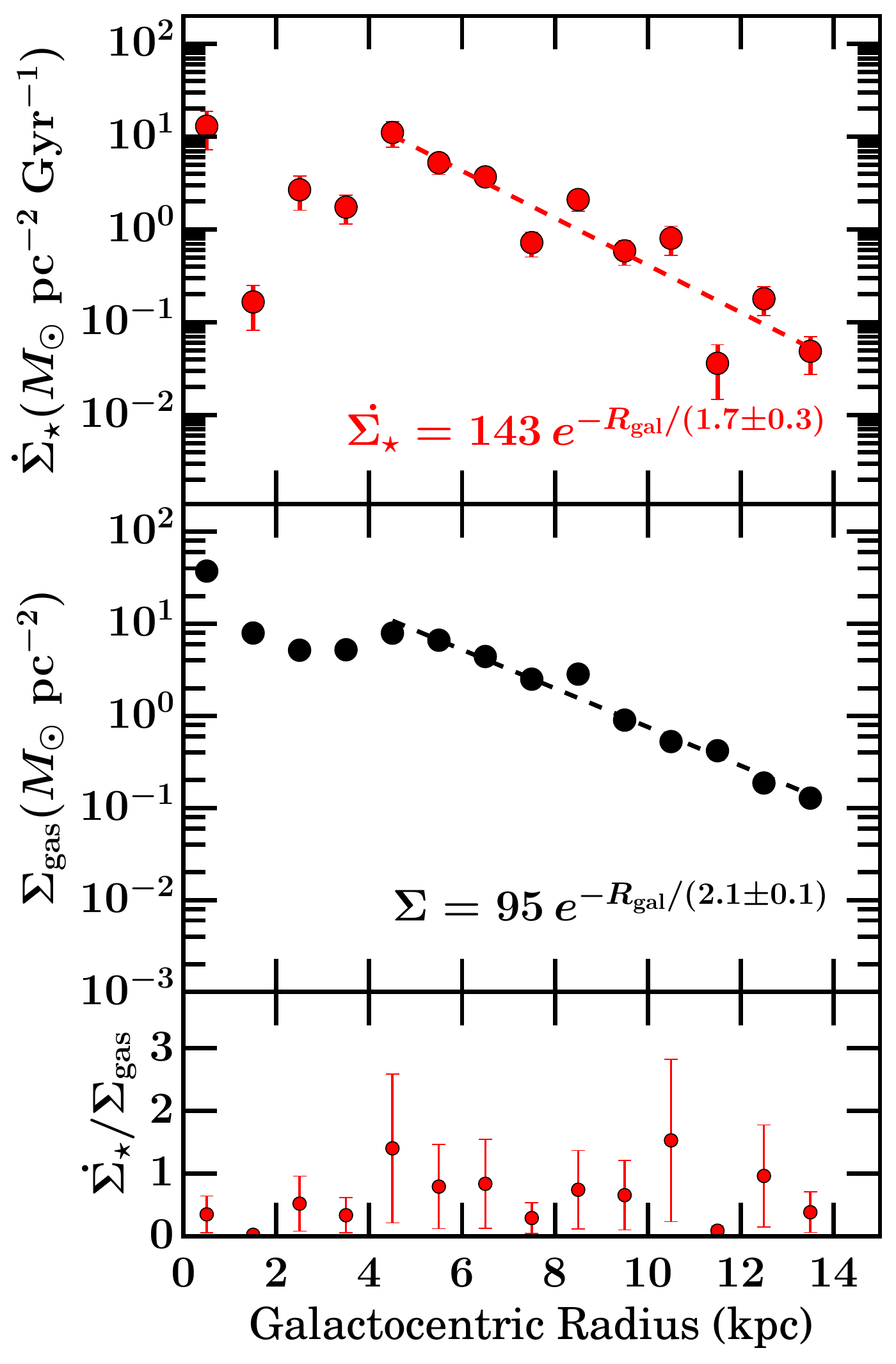}
\caption{Top: star formation rate per unit 
area $\dot\Sigma_\star$ as a function of galactocentric radius. 
Middle: gas surface density profile $\Sigma_{\rm gas}$.
Bottom: the ratio between the two profiles $\dot\Sigma_\star/\Sigma_{\rm gas}$.
Star formation is observed to follow the gas: they all show similar scale radius. 
Both profiles show a break at $\sim4\kpc$, consistent 
with the presence of a molecular ring.}
\label{fig12}
\end{figure}

\section{THE MILKY WAY STAR FORMATION RATE SURFACE DENSITY PROFILE}
\label{sec:sigma}

\citet{kennicutt12}, in their Figure 7, show that the star formation rate per unit area 
$\dot\Sigma_\star$ falls off more slowly than the surface density of molecular gas 
$\Sigma_{\rm gas}$. By contrast, in the nearby galaxy NGC 6946, 
$\dot\Sigma_\star$ is observed to follow $\Sigma_{\rm gas}$ closely.
The information on star formation rate used by \citet{kennicutt12} 
date back to \citet{guesten82}. 
The catalog of SFCs and MML16 clouds used in this paper should 
provide the most up-to-date and the most 
complete estimate of star formation rate and the molecular gas mass.
Using the sample of star-forming clouds, we illustrate in 
Figure \ref{fig12} how even in the 
Milky Way, $\dot\Sigma_\star$ tracks well $\Sigma_{\rm gas}$
as a function of galactocentric radius.
In particular, both the star formation rate and the gas density profiles 
are well-fitted with an exponential profile of scale length 
$\sim$2.0. The trend observed in $\dot\Sigma_\star$ is anti-correlated 
with the profile of $\alpha_{\rm vir}$ shown 
in Figure 14 of MML16: where $\alpha_{\rm vir}$ rises, $\dot\Sigma_\star$ 
dips (inside 4 $\kpc$ and outside 8 $\kpc$) and where $\alpha_{\rm vir}$ 
dips, $\dot\Sigma_\star$ rises (between 4 and 8 $\kpc$). This anti-correlation 
is expected since stars should form in gravitationally-bound regions. 

\section{DISCUSSION AND SUMMARY}
\label{sec:disc}

We have demonstrated in this paper the importance of the scatter 
in the distribution of $\epsilon_{\rm ff}$ and 
$\epsilon$. Any successful model of cloud-scale star formation should 
reproduce not only the mean star formation rate but also
the large scatter about the median.
The spread in $\epsilon$ has long been known to be enormous: 
\citet{mooney88}, \citet{scoville89}, and \citet{mead90} find a 
minimum-maximum range of three orders of magnitude in $\epsilon$ as 
probed by FIR luminosity in a small sample of nearby molecular clouds. 
More recently, \citet{lada10} and \citet{heiderman10} count young stellar 
objects in a sample of nearby clouds, and show that the values 
of $\epsilon_{\rm ff}$ range up to a factor of ten larger than
the nominal value of $0.02$. \citet{evans14} report approximately 
an order of magnitude spread in $\epsilon_{\rm ff}$ both above 
and below 0.01 in nearby molecular clouds from the c2d and Gould's belt survey. 
Using dense clumps from ATLASGAL, \citet{heyer16} find $\epsilon_{\rm ff}$ 
can be as low as $\sim$0.001.

We combined the Milky Way all-sky catalog of GMCs by 
M-A., Miville-Desch\^enes et al. (2016) and 
an all-sky catalog of SFCs by \citet{lee12} to build a large collection 
of star-forming clouds (see Section \ref{sec:sfgmc}). 
Our MML16 catalog of GMCs contains 
5469 valid clouds, which is about an order of magnitude more clouds
than all previously published catalogs. 
Using the statistical power afforded by this catalog, we showed that 
both $\epsilon$ and $\epsilon_{\rm ff}$ of star-forming GMCs 
in the Milky Way have large dispersions:
$\sigma_{\log\epsilon,}=0.79\pm 0.22$ dex,
and $\sigma_{\log\epsilon_{\rm ff}}=0.91\pm0.22$ dex, 
(see Sections \ref{sec:sfe} and \ref{sec:sfr_ff}). 
These results confirm and extend earlier observations. 
The error in the scatter is dominated by 
the statistical variations in the IMF 
(see Section \ref{sec:flux}).

Variations in internal cloud properties or their large scale motions 
cannot account for the large scatter in 
$\epsilon_{\rm ff}$. 
We rule out the constant star-formation rate model of \citet{krumholz05} 
since cloud-to-cloud variations in the virial parameter $\alpha_{\rm vir}$ or 
the Mach number ${\cal M}$ do not produce a large enough scatter 
in $\epsilon_{\rm ff}$ (see Section \ref{ssec:turbsf}). 
Even the improved model of turbulence-regulated star 
formation by \citet{hennebelle11}---that has a strong 
dependence on the Mach number---cannot account for the 
observed scatter.
Similarly, we find at least an order of magnitude scatter in star formation rate 
about the collision-induced star formation law proposed by \citet{tan00};
the rate at which gas is converted to star on a cloud-to-cloud 
collisional timescale cannot be a constant (see Section \ref{ssec:collision}).
We see some evidence of decreasing 
$\epsilon_{\rm ff}$ with larger $\tau_{\rm ff}/\tau_{\rm dyn}$ 
as expected by \citet{padoan12} 
but the large scatter for a given time ratio cannot be explained by 
varying magnetic field intensities (see Section \ref{ssec:sfr-vir}). 

One way to produce a large dispersion in both 
$\epsilon$ and $\epsilon_{\rm ff}$ is to arrange for a 
time-variable rate of star formation in a GMC with fixed gas mass, density, 
and velocity dispersion (see Section \ref{ssec:time_dependent}).
Generalizing the turbulence-regulated star formation model of \citet{krumholz05} 
to allow time-variable $\epsilon_{\rm ff}$ in the prescription of \citet{feldmann11}, 
we find that $\epsilon_{\rm ff} \propto t^2$ is most 
consistent with our data.
We conclude that star formation is dynamic on the 
GMC-scale.\footnote{The time-varying $\epsilon_{\rm ff}$ may arise from 
the evolution of star-forming clumps embedded in each GMC.}

Our study concerns the properties of star formation in clouds 
that have already formed or in the process of forming stars. 
By construction, we have limited 
our analysis to 191 SFC-GMC complexes out of 5469. 
We found 1014 clouds unmatched to SFCs that have the potential 
to form star clusters (see Section \ref{ssec:lifetime}).
We surmise that most (5469 - 1014 - 191 = 4264; 78\% by number) of the GMCs 
are gravitationally unbound 
and will never form massive star clusters.

It is possible that some locally 
dense clumps within an unbound GMC form 
stars \citep[see e.g.,][]{dobbs11}. These clumps likely 
give birth to lower mass star clusters 
that our star formation tracers are not sensitive to. 
Throughout our analysis, we used free-free emission 
as a proxy for stellar mass. A more direct measurement would be to count
young stellar objects (YSOs). For known massive clusters, 
we find a generally good agreement in the measured stellar mass 
between free-free emission and YSO counts (see Figure \ref{fig2}).
A more complete comparison may be possible 
using all-sky YSO catalogs \citep[e.g.,][]{marton16}. This is an important 
and natural avenue for future improvement. 

\subsection{Comments on Stellar Feedback}
\label{ssec:feedback}
If the rate of star formation in the most massive 
($10^6M_\odot$ and higher) GMCs does accelerate, 
the process of star formation must halt before $\epsilon\gtrsim 0.1$, 
roughly the largest value we see in such clouds. 
Stellar feedback---in the form of stellar winds, radiation pressure, protostellar jets, 
and supernovae---from massive star clusters can disrupt the natal cloud and 
inhibit future star formation. 
The idea that stellar feedback destroys GMCs was proposed long ago by \citet{larson81}. 
Galactic-scale numerical simulations (e.g., \citealt{hopkins11} and \citealt{fg13}) 
find that the effects of stellar feedback are necessary for regulating star 
formation rates to the Kennicutt-Schmidt value. Turbulence in the interstellar medium 
alone cannot slow the rate of star formation.

The existence of large expanding bubbles---a few to $\sim$100 parsecs wide---associated with 
regions bright in free-free emission is evidence for the effects of stellar feedback 
from young clusters (referred to as SFCs by \citealt{rahman10} and \citealt{lee12}). 
These massive young clusters, which power 
most of the free-free emission in the Milky Way, are identified by clear cavities in 
the 8$\mu$m emission. The clusters are found in massive GMCs, but the  
star clusters are not enshrouded in molecular gas---they are instead 
in regions of ionized gas which are enshrouded in molecular gas. 
In some cases, the massive stars may still be accreting gas from their natal protostellar 
disks but the disks are not accreting gas from the host GMC. 

We have examined our sample to see if the feedback from stars 
is affecting their host GMCs globally. 
There was no clear correlation between the variations in $\alpha_{\rm vir}$ 
and the ratio of disruptive forces (gas and radiation pressure) to the 
binding force (dynamical pressure) of the host GMC. We also found no clear 
correlation between the size of the bubbles blown by SFCs in the host clouds 
and $\alpha_{\rm vir}$.
It may be that the effects of radiation and gas pressure will only become evident 
later in the evolutionary history of the GMC, or it may be that other forms 
of feedback (e.g., supernovae) are responsible for disrupting the 
clouds. Given that GMCs in the Milky Way are found near spiral arms, combined with our finding 
that the a large fraction of $10^6M_\odot$ GMCs are most likely gravitationally
bound ($\alpha_{\rm vir} < 1$), it seems unlikely that they simply disperse on their own.

\acknowledgments
The creation of the MML16 catalog of GMCs is made possible by 
the all-sky CO data provided by T. Dame.
We thank the anonymous referee for their constructive feedback 
that helped to improve this paper significantly.
We thank L. Blitz, G. Chabrier, E. Chiang, C-A. Faucher-Gigu\'{e}re, C. Federrath, 
R. Feldmann, N. Gnedin, P. Hopkins, M. Krumholz, P-S. Li, C. Matzner, C. Mckee, 
E. Ostriker, E. Quataert,  S. Stahler, J. Tan, 
and E. Vazquez-Semadeni for valuable discussions. 
We also thank H. Isaacson, R. Trainor, 
and J. Wang for their advice in statistical analyses. 
EJL is supported in part by NSERC of Canada under PGS D3 and the Berkeley fellowship.
NWM is supported in part by NSERC of Canada. This research was undertaken, in part, 
thanks to funding from the Canada Research Chairs program.

\bibliography{gmc}

\end{document}